\documentclass[prb,showkeys,showpacs]{revtex4-1}
\usepackage{graphicx}
\usepackage{natbib}
\usepackage{bm}
\usepackage{amssymb}
\usepackage{amsmath}
\usepackage{euscript}
\usepackage{mathrsfs}
\usepackage{esint}
\usepackage{slashed}

\begin{document}

\title{Premixed flame propagation in vertical tubes}

\author{Kirill A. Kazakov}

\affiliation{Department of Theoretical Physics, Physics Faculty, Moscow State University, 119991, Moscow, Russian Federation}

\begin{abstract}
Analytical treatment of premixed flame propagation in vertical tubes with smooth walls is given. Using the on-shell flame description, equations describing quasi-steady flame with a small but finite front thickness are obtained and solved numerically. It is found that near the limits of inflammability, solutions describing upward flame propagation come in pairs having close propagation speeds, and that the effect of gravity is to reverse the burnt gas velocity profile generated by the flame. On the basis of these results, a theory of partial flame propagation driven by the gravitational field is developed. A complete explanation is given of the intricate observed behavior of limit flames, including dependence of the inflammability range on the size of the combustion domain, the large distances of partial flame propagation, and the progression of flame extinction. The role of the finite front-thickness effects is discussed in detail. Also, various mechanisms governing flame acceleration in smooth tubes are identified. Acceleration of methane-air flames in open tubes is shown to be a combined effect of the hydrostatic pressure difference produced by the ambient cold air and the difference of dynamic gas pressure at the tube ends. On the other hand, a strong spontaneous acceleration of the fast methane-oxygen flames at the initial stage of their evolution in open-closed tubes is conditioned by metastability of the quasi-steady propagation regimes. An extensive comparison of the obtained results with the experimental data is made.
\end{abstract}
\maketitle

\section{Introduction}

Flame propagation in vertical tubes takes a special place among classical problems of premixed combustion. Being one the simplest and practically important settings, vertical tubes were chosen to define a standard of mixture inflammability.\cite{cowardjones1952,zabetakis} Since determination of the inflammability limits -- the endpoints of the range of fuel concentrations over which a given mixture is able to sustain laminar flame propagation -- is of particular interest in regard to combustion safety, much effort has been spent on their experimental and theoretical study. \cite{cowardjones1952,zabetakis,levy1965,lovachev1971,bregeon,jarosinski,buckmaster,vonlavante} The inflammability limits are measured in vertical tubes sufficiently wide to make negligible heat losses to the tube walls, and sufficiently long to let the initial disturbance produced by the ignition die out. This elimination of the external energy factors is meant to identify inflammability as an intrinsic characteristic of the mixture. It is well-known, however, that dynamics of virtually all laboratory flames is strongly affected by the terrestrial gravitational field, the gravity effect being especially pronounced near the inflammability limits where the normal flame speed is lowest. The comparatively narrow practical question of measuring inflammability limits has thus grown into an extensive area of research of gravity-driven flame evolution under the critical conditions of extinction. Still, despite considerable experimental data collected, there has been little progress on the theoretical side of the problem. This is primarily because the usual complication of the flame theory -- essential nonlinearity of the gas flows induced by the flame -- becomes extreme under the most common laboratory conditions. In fact, a simple dimensional analysis suggests that the relative gravity impact can be quantified by the Froude number,
$$(acceleration~of~gravity)\times  (characteristic~length)/(normal~flame~speed)^2.$$
It follows that the global structure of all hydrocarbon-air flames, in particular, their propagation speed is strongly affected by the gravitational field in tubes wider than 1\,cm. Furthermore, in the case of limit flames, gravity shows itself not only at the hydrodynamic scale, but also at all scales down to the smallest -- the flame front thickness. For instance, the front thickness of a limit methane-air flame in a $5$\,cm-diameter tube is $l_f\approx 0.2$\,mm, while its normal speed $U_f\approx 5$\,cm/s, for which the above dimensionless ratio is about unity. In view of such a wide scale separation, it is not surprising that this problem still remains beyond the capabilities of direct numerical methods.

As to the analytical approach, there have been several attempts to quantitatively describe the nonlinear flame stabilization by the gravitational field. The central question here is the way the flow nonlinearity is dealt with. Within the traditional approach based on explicit solving of the hydrodynamic equations for the bulk flows, there are essentially two ways: either to treat the nonlinearity perturbatively, or to introduce a suitable ansatz to model the flow structure. The first option, realized in Ref.~\cite{sivashinsky}, is the classical approximation of small density contrast of fresh and burnt gases, which until recently has been the only self-consistent framework to treat flame nonlinearities. However, general applicability of the results of this analysis to real flames is restricted by the fact that in practice, the fresh-to-burnt gas density ratio, $\theta,$ significantly exceeds unity (typical values are 5 to 10). In the present context, this applicability is further limited by the requirement of smallness of the Froude number, because for flames with $(\theta-1) = O(1)$ the gravity-induced flow nonlinearity becomes strong as this number exceeds unity, which has already been identified as a common condition. A notable example of the other traditional approach to the problem is the unimodal model\cite{pelce} that uses a single-mode approximation of the flow upstream the flame front to complete the system of Bernoulli integrals and mass- and momentum-conservation laws. Yet, it is clear in advance that employing one mode, or a few, will not suffice to approximate a strongly nonlinear flow, and this insufficiency shows itself in this model as a loss of nontrivial solutions at large Froude numbers.

It is only recently that the strongly nonlinear flames became accessible to theoretical analysis, with the invention of the on-shell flame description.\cite{kazakov1,kazakov2} This approach circumvents the nonlinearity problem through the use of dispersion relations for the up- and downstream gas flows. Its main result -- the so-called master equation -- is an exact consequence of the fundamental flow equations, and as such is very complicated in general. But it dramatically simplifies for flames with elongated fronts, as is often the case with flames in the presence of strong gravity. This simplification was recently employed to study flame propagation in horizontal tubes within the zero front-thickness approximation.\cite{kazakov3,kazakov4} The purpose of the present paper is to investigate vertical flame propagation using the same means. In carrying out this investigation, however, one faces the necessity to take into account finiteness of the flame front thickness. From a purely formal standpoint, vertically propagating flames can be consistently treated within the zero-thickness approximation. Moreover, the results thus obtained do find application to real flames, but it turns out impossible to fully understand the observed flame behavior neglecting the finite front-thickness effects completely. This is in contrast to the horizontal case where these effects may or may not be noticeable, but they are never so large as to, say, qualitatively change dependence of the flame propagation speed on the fuel concentration, or even preclude existence of some propagation regimes. The need to include the finite front-thickness corrections raises in turn the old question of their structure. Despite several decades of investigation, this is still a matter of debate. By this reason, after recalling the basic results of the on-shell description in Sec.~\ref{onshell}, this question is addressed in Sec.~\ref{ffthickness}. It is proved that the ambiguity in positioning a discontinuity surface that replaces the flame front, which is the stumbling block of the conventional treatments, can be resolved by considering evolution of the individual gas elements, rather than the flow velocity fields. In Sec.~\ref{mainequations}, the master equation for finite front-thickness flames subjected to a strong gravitational field is reduced to a system of ordinary differential equations and boundary conditions. This system is then used in Sec.~\ref{applications} to study flame dynamics near the limits of inflammability and acceleration of upward propagating flames. The general procedure for calculating finite front-thickness corrections -- an asymptotic matching of the inner and outer solutions of gasdynamic equations -- implies smallness of the corresponding terms: corrections to the normal front speed and other flow variables must be sufficiently small compared to their values for a zero-thickness flame. This requirement brings its usual specifics into applications of the asymptotic analysis, primarily that the range of applicability of the theory is not known {\it a priori}, because no accurate analytical estimate exists of the remainder of the asymptotic series. This series diverges already from the first-order term whenever its magnitude becomes comparable to the zero-thickness value. As mentioned above, this happens with the methane-air flames near the lean inflammability limit in a $5$\,cm-diameter tube, and we will see in Sec.~\ref{critical} that the small-$l_f$ asymptotic expansion blows up in this case indeed. Since the range of applicability of the asymptotic analysis in each case is to be determined empirically, it is important to determine {\it initial trends} in the flame behavior with respect to the asymptotic expansion parameter. By this reason, applications of the developed theory begin with comparison of the experimental results with the predictions of the zero-thickness theory. ``Switching on'' the finite front-thickness effects then allows one to see clearly the difference they produce in the flame structure. This is particularly important in the theory of partial flame propagation and extinction, developed in Sec.~\ref{inflammability}. Numerical analysis reveals quite nontrivial properties of the solutions describing near-limit flames. On the basis of these results, a complete explanation is given of the intricate observed behavior of these flames. Comparison of the numerical solutions with the experimental data on flame propagation in open and semi-open tubes also helps identify various mechanisms governing flame acceleration in smooth tubes, which is done in Sec.~\ref{acceleraion}. The main results of the present paper are summarized and further discussed in Sec.~\ref{conclusions}, which also outlines related open issues and prospects for future work.

\section{On-shell equations}\label{onshell}

\begin{figure}
\includegraphics[width=0.35\textwidth]{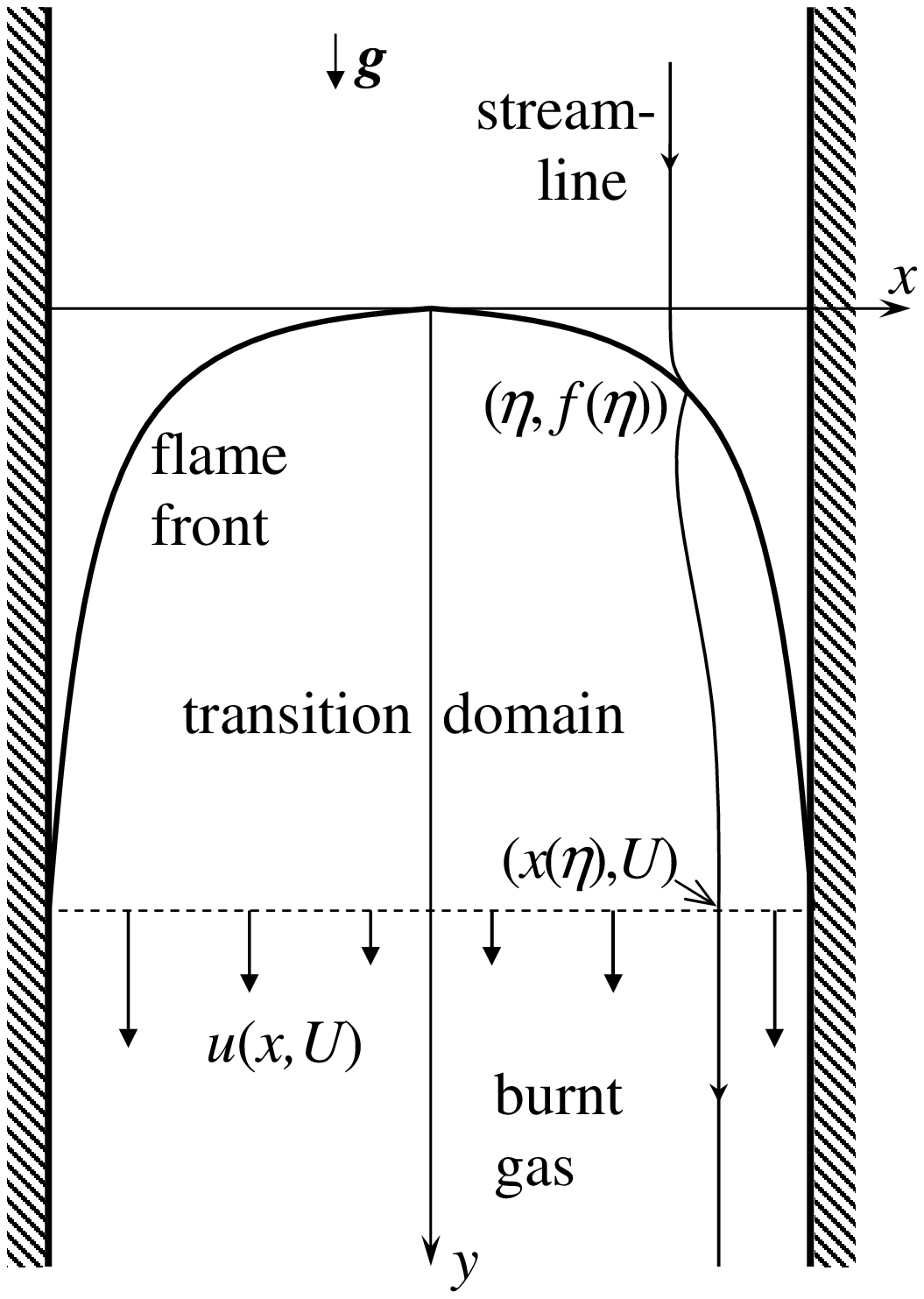}
\caption{Schematics of upward flame propagation. Vertical arrows attached to the lower boundary (dashed line) of the transition domain $y\in (0,U)$ depict burnt gas velocity profile generated by the flame in the case $g\gg U^2_f/b$.}\label{fig1}
\end{figure}

Consider flame propagation in an initially quiescent gaseous mixture filling vertical channel of width $b.$ Experiments with flames in tubes show that soon after ignition at an open tube end, the flame front assumes a characteristic shape depicted schematically in Fig.~\ref{fig1}. The subsequent flame evolution depends essentially on whether the other end is open or closed. In the latter case, the flame moves steadily over a considerable distance (the so-called uniform movement) until the flame-acoustic interaction comes into play. On the other hand, if the upper tube end is open, steady propagation regimes no longer exist, as the flame will accelerate. Yet, the acceleration will be small in a sufficiently long tube, so that the flame evolution can be considered quasi-steady in the rest frame of the flame, the effect of acceleration being described as that of an additional uniform gravitational field equal in value and opposite to the flame acceleration in the laboratory frame.\cite{kazakov4} In any case, therefore, we can choose the reference frame attached to the steady flame, with the origin at its tip, denoting $x$ the horizonal coordinate, and $y$ the coordinate directed downward along the channel centerline. The front shape of vertically propagating flames is often symmetric with respect to the line $x=0.$ Accordingly, lengths will be measured in units of the channel width $b$ or half-width $b/2,$ in the asymmetric and symmetric cases respectively. The normal speed of planar flame relative to the fresh mixture, $U_f,$ will be taken as the unit of gas velocity $\bm{v} = (w,u).$ It is occasionally convenient, especially in applications, to switch from these natural units back to the $SI$-units; the latter will always be specified explicitly, while dimensionless figures will stand for quantities measured in the natural units.

Since $U_f$ is much smaller than the sound speed, the gas flow induced by the flame can be treated as incompressible. Then the gas-velocity distributions along the flame front and the front position $y=f(x)$ satisfy the following complex integro-differential equation (the master equation),\cite{kazakov1,kazakov2,jerk1,jerk2} which is an exact consequence of the fundamental gasdynamic equations for ideal flow
\begin{eqnarray}\label{master1}&&
2\omega_-' + \left(1 +
i\hat{\EuScript{H}\,}\right)\left\{[\omega]' -
\frac{M\omega_+}{v^2_+} +\frac{1+if'}{2}\int_{-1}^{+1}d\eta\frac{M(\eta)\omega_+(\eta)}{v^2_+(\eta)}\right\} = 0\,,
\end{eqnarray}
\noindent where $\omega = u + iw$ is the complex velocity, $[\omega] = \omega_+ - \omega_-$ its jump across the front; prime denotes $x$-differentiation, the subscript $- (+)$ restriction to the front of a function defined upstream (downstream) of the front, {\it e.g.,} $w_-(x) = w(x,f(x)-0)$; $N = \sqrt{1 + f'^2},$ $v^n$ is the normal gas velocity (the normal $\bm{n}$ to the front points to the burnt gas), $\sigma=\partial u/\partial x - \partial w/\partial y$ is the vorticity; $M \equiv Nv^n_+\sigma_+$ is the so-called memory kernel; finally, the $\hat{\EuScript{H}\,}$-operator is defined by
\begin{eqnarray}
\left(\hat{\EuScript{H}\,}a\right)(x) = \frac{1 + i
f'(x)}{2}~\fint_{-1}^{+1}
d\eta~a(\eta)\cot\left\{\frac{\pi}{2}(\eta - x +
i[f(\eta) - f(x)])\right\}\,,\nonumber
\end{eqnarray}
\noindent where the slash denotes the principal value of the integral. Equation~(\ref{master1}) is to be complemented by a relation defining the local burning rate (the evolution equation). This relation is of the form
\begin{eqnarray}\label{evolution}
v^n_- = 1 - S(f',w_-,u_-),
\end{eqnarray}
\noindent where $S(f',w_-,u_-)$ is a quasi-local functional of its arguments that describes the effect of transport processes inside the flame front on the normal front speed. The on-shell vorticity $\sigma_+$ and the velocity jump $[\omega]$ are similar functionals whose structure is discussed in the next section.

\section{Finite front-thickness effects}\label{ffthickness}

\subsection{Flame as a surface of discontinuity}\label{discontinuity}

The structure of finite front-thickness corrections to the normal flame speed and gas velocity jumps at the front is probably the most controversial topic in the premixed flame theory. Though the general framework for calculating these corrections was developed more than thirty years ago, its results are still a matter of hot debate. The reason is an apparent ambiguity in the definition of the discontinuity
surface which replaces the actual thick flame front, and which the symbols $\pm$ in the above equations refer to. To be specific,
in the first order with respect to the small front thickness $l_f$
(measured in units of $b$ or $b/2,$ depending on the flame symmetry),
the functional $S$ reads
\begin{eqnarray}\label{sfunctional}
S(f',w_-,u_-) = \mathscr{L}_s\frac{(v^{\tau}_-)'}{N} + \mathscr{L}_c\left(\frac{f'}{N}\right)',
\end{eqnarray}
\noindent where $v^{\tau}_- = (w_- + f'u_-)/N$ is the tangential to the front component of the fresh gas velocity, and $\mathscr{L}_s,\mathscr{L}_c = O(l_f)$ are length parameters (Markstein lengths) quantifying the flame stretch and front curvature effects on the flame speed.\cite{mikhelson1890,markstein1951,markstein1964,joulin1983,clavin1985,joulin1989} (Flame stretch, $(v^{\tau}_-)'/N,$ includes the curvature effect on its own, as it can be written as the difference of a term proportional to the front curvature and the normal-normal component of the flow strain, Cf. Eq.~(\ref{displacement}) below.) Both the mathematical definition of these parameters and their experimental determination are seemingly problematic in view of the fact that displacing the discontinuity surface within the flame front thickness for a distance $\delta \xi$ along the normals changes the left hand side of Eq.~(\ref{evolution}) by
\begin{eqnarray}\label{displacement}
\delta \xi\left(\frac{\partial v^n}{\partial n}\right)_- = \delta \xi\left[v^n_-\left(\frac{f'}{N}\right)' - \frac{(v^{\tau}_-)'}{N}\right].
\end{eqnarray}
\noindent The latter equality (which is a direct consequence of the flow incompressibility) shows that the displacement adds to $\mathscr{L}_s,\mathscr{L}_c$ terms $O(l_f)$ that are of the same order of magnitude as the Markstein lengths themselves.\cite{joulin1989,bechtold,clavin2011} Opinions regarding this state of affairs vary from assertions that such an ambiguity is in the nature of things and can be used to optimize the first-order approximation, to the statement that it is merely an indication on the
impossibility to obtain a correct description of a finite-thickness flame by replacing it with a discontinuity surface. The subsequent analysis leads to a conclusion lying in between these two extreme standpoints.

A prevailing opinion is that the freedom in locating the discontinuity surface within the flame is real, and each particular choice is basically as legitimate as any other, the preference of one over the others being purely a matter of convenience, while final results for the flame dynamics are independent of this choice.\cite{clavin1985,class2003,clavin2011,matalon} In this connection, an analogy with the thermodynamic problem of equilibrium of two heterogeneous substances is sometimes invoked to stress generality of this kind of arbitrariness. As this analogy helps identify an important ingredient clarifying the issue, the classical Gibbs' analysis\cite{gibbs1885} of the latter problem will be now briefly recalled.

Consider an equilibrium state of a heterogeneous system consisting of two masses each of which is homogeneous in its composition. In the absence of a contact, each of the masses would have homogeneous not only its composition, but also the energy density. The latter becomes inhomogeneous as a result of interaction of the two substances. Because, however, the molecular forces are short-ranged, this inhomogeneity is confined in a very thin layer surrounding the contact region. This fact offers a simple method to take into account the effect of inhomogeneity, that is of the mass contact, on the system energy. Namely, one introduces a mathematical surface separating the masses, drawn somewhere within the inhomogeneity layer, and replaces the actual matter distribution with a fictitious one, which is assumed to be completely homogeneous in each of the masses quite up to the separating surface. Then according to the first law of thermodynamics, the energy difference of the original and fictitious systems, called naturally the surface energy $\varepsilon^s,$ must be of the form
$$\delta\varepsilon^s = \Sigma \delta s + C_1\delta c_1 + C_2\delta c_2 + \dots,$$ where $\delta s$ is a variation of the surface area, $\delta c_{1,2}$ variations of its principal curvatures, $\Sigma, C_{1,2}$ are some functions of the thermodynamic state of the system and the {\it separating surface position}, and dots stand for the contributions due to variations of the system entropy {\it etc.} Of course, $\Sigma, C_{1,2}$ depend on the position of the separating surface because so does the energy of the fictitious system. But the energy of the system in question, which is the sum of the fictitious system energy and surface energy, does not. A simple geometrical consideration shows that a shift of this surface induces variations of $C_{1,2}$ which are proportional to $\Sigma$ and the shift distance. By estimating various terms in the above equation, Gibbs then argues that it should always be possible to place the surface within the inhomogeneity layer so as to eliminate the terms involving curvature variations. The quantity $\Sigma$ defined this way is called surface tension. The system energy is thus represented in a very simple and convenient form -- it is the sum of energies of homogeneous masses unaffected by the contact, and of the surface energy controlled by a single parameter which can be measured directly with the help of the relation
\begin{eqnarray}\label{ptension}
[p] = \Sigma (c_1+c_2),
\end{eqnarray}
\noindent
where $[p]$ denotes the pressure jump at the surface. It is to be stressed that there is no ambiguity in the definition of the surface tension whatsoever. This is because in principle, the system energy can always be measured to any desired accuracy. Likewise, the energy of the fictitious system can also be thought of as known exactly, because its density is uniform over each of the two masses (as are all other thermodynamic quantities), and coincides with the respective bulk value in the original system. Therefore, for each given position of the separating surface, the quantities $\Sigma, C_{1,2}$ as defined via the difference of these two energies, all have well-defined values. In particular, their accuracy has nothing to do with the assumed thinness of the inhomogeneity layer. Still, although one can, in principle, follow the changes in $\Sigma, C_{1,2}$ as the separating surface takes on different positions within the inhomogeneity layer, this is practically difficult, and its particular position securing disappearance of the curvature variations from $\delta\varepsilon^s$ is left practically undetermined. It is here that the thinness of the inhomogeneity layer is used to neglect this ambiguity, and this is also the source of experimental uncertainty in the value of the surface tension. In fact, shifting the surface within the inhomogeneous layer produces a relative change in the surface curvature of the order of the layer thickness times the surface curvature, hence a change of the same order in $\Sigma,$ by virtue of the relation (\ref{ptension}).

Turning back to flames, the situation is very similar except that the set of physical quantities is now a continuum. Namely, the motion of fresh and burnt gases is what we are interested in, and the problem is to determine approximately the effect of transport processes inside the flame front on the bulk gas motion using the smallness of the flame front thickness. Thus, the set of bulk trajectories of gas elements is the counterpart of energy of the heterogeneous system considered above, whereas the normal flame speed and the jumps of flow variables at the front are counterparts of $\Sigma, C_{1,2}.$ This time our aim is to capture effects related to the finite front thickness, that is, we will have to control position of the discontinuity surface within the flame front. This fictitious surface is the analog of the separating surface between the masses in contact, and is introduced in order to facilitate the study of the bulk gas dynamics. The density of gases in the upstream and downstream regions of this surface is assumed to be uniform quite up to the surface, and to coincide with the bulk density of fresh and burnt gases, respectively. Just as the energy of the two-mass system was not to be affected by manipulations with the inhomogeneous layer, we have to require that the bulk properties of gases be unaffected by the replacement of the finite-thickness flame front with a surface of discontinuity. We thus impose the following condition:

(C) {\it Positions of the gas elements which are remote from the flame front at any given instant must be left unaffected by the replacement of the flame front with a discontinuity surface.}

Evidently, this is the most general criterion of validity of any flame front model. The term remote means that the gas density gradient at the given position is negligibly small; since the gas temperature varies exponentially within the front, this requirement is practically satisfied at distances of a few thermal lengths (defined as $\chi/U_f,$ $\chi$ denoting the thermal diffusivity of gases) away from the surface of maximal reaction rate.

We will now derive an analytical expression of this condition. The subsequent consideration is quite general, and is not limited by the assumptions of two-dimensionality or steadiness of the flow. To begin with, an important remark on the nature of physical quantities we deal with is to be made. The point is that although these quantities (positions of the gas elements, or their trajectories, in case the processes is considered in time) pertain to the bulk regions, they are nevertheless directly affected by the processes inside the flame front. In fact, it is a characteristic property of flames which distinguishes them, say, from bubbles, that sooner or later, each gas element crosses the front. In this sense, the whole burnt gas flow is dependent on the front structure, and part of the condition (C) is that replacing the front by a discontinuity surface must leave positions of the burnt gas elements intact. Take any gas element and consider part of its trajectory, $\bm{R}(t),$ between a point $\bm{R}_1 = \bm{R}(t_1)$ belonging to the bulk of fresh gas and a point $\bm{R}_2 = \bm{R}(t_2)$ in the bulk of burnt gas (capital letters are used to distinguish quantities related to the original system, that is before the replacement). In the picture where the front is replaced by a discontinuity surface, trajectory $\bm{r}(t)$ of this element runs over the same set of points as $\bm{R}(t),$ by virtue of (C), except probably in the layer of thickness $O(l_f)$ surrounding the surface. But in order for (C) to be satisfied, it is also necessary that the travel time of the element in the two pictures be the same. Otherwise, even if the trajectories $\bm{R}(t)$ and $\bm{r}(t)$ were coincident in the bulk upstream of the front, a lag between them would necessarily appear downstream. Writing the element of travel time as $|d\bm{R}|/|\bm{V}|,$ this condition reads
\begin{eqnarray}\label{traveltime}
\int_{\bm{R}_1}^{\bm{R}_2}\frac{|d\bm{R}|}{|\bm{V}|} = \int_{\bm{R}_1}^{\bm{R}_2}\frac{|d\bm{r}|}{|\bm{v}|}\,.
\end{eqnarray}
\noindent Components of the auxiliary velocity field $\bm{v}$ have finite jumps at the discontinuity surface, whereas the true gas velocity $\bm{V}$ is independent of its position. It is clear, therefore, that the solution to Eq.~(\ref{traveltime}), considered as an equation for this position, is unique. We thus arrive at the conclusion that {\it there is actually no freedom in placing the discontinuity surface within the flame front}, in contrast to the problem of calculating the energy of two masses in contact.

Now that uniqueness of the solution has been established, it remains to be demonstrated that placing the discontinuity surface as dictated by Eq.~(\ref{traveltime}) it is possible to construct a correct flame description. Specifically, one has to show that solutions of the hydrodynamic equations for a fluid of constant density exist on each side of the surface, such that the condition (C) is met. It is to be noted that this task is not that trivial as it might seem at the first sight. Replacement of the front of finite thickness $l_f$ by a surface separating gases of uniform density does change the flow at distances $O(l_f)$ from the surface. In view of the essential non-locality of hydrodynamics involved in the process, this generally results in a $O(l_f)$ relative change of the flow variables in the bulk. That (C) can be satisfied nevertheless will be shown in the first order approximation with respect to the front thickness. For this purpose, Eq.~(\ref{traveltime}) will be simplified with the same accuracy. Since the integrands in this equation differ only at distances $O(l_f)$ from the discontinuity surface, the difference of the two integrals in Eq.~(\ref{traveltime}) is already $O(l_f).$ By the same reason, the substance of Eq.~(\ref{traveltime}) is independent of $\bm{R}_1, \bm{R}_2,$ provided that they are both in the respective bulk regions. Therefore, velocity gradients in these regions can be neglected in the same approximation. In these circumstances, it is convenient to go over to the reference frame where the discontinuity surface is at rest at the point of its intersection with $\bm{r}(t).$ Since the variations of $\bm{V}$ and $\bm{v}$ at distances $O(l_f)$ in the tangential directions are $O(l_f),$ these velocities can be evaluated along the projections of $\bm{R}(t)$ and $\bm{r}(t),$ respectively, onto the same normal to the surface at the point of intersection. Both integrals in Eq.~(\ref{traveltime}) are thus expressed in terms of the same coordinate, $\xi,$ along the normal at the point of intersection, and we obtain
\begin{eqnarray}\label{traveltimef}
\int_{-\infty}^{+\infty}d\xi\left(\frac{1}{V^n} - \frac{1}{v^n}\right) = 0,
\end{eqnarray}
\noindent where $\pm\infty$ mean that the integration limits are sufficiently far from the surface. Since integration in this formula is effectively over an interval of length $O(l_f),$ $v^n$ in the integrand is to be evaluated in the zero-order approximation in the front thickness, that is $v^n = 1$ upstream the discontinuity surface, and $v^n = \theta$ downstream.

To prove that (C) can be satisfied, let us follow evolution of any gas element as it passes through the flame. Since the flow equations governing bulk dynamics of gases in the picture where the front is replaced with a discontinuity surface are identical with those of the original system, trajectories of the gas element in the bulk of fresh gas in the two pictures are also identical. Their subsequent behavior as the element approaches the flame can be found by integrating the relations
\begin{eqnarray}\label{trajectories}
\frac{d\bm{R}}{dt} = \bm{V}(\bm{R},t), \quad \frac{d\bm{r}}{dt} = \bm{v}(\bm{r},t),
\end{eqnarray}
\noindent where the true velocity field $\bm{V}$ is, in principle, known exactly, whereas the fictitious field $\bm{v}$ is to be found as a continuation of $\bm{V}$ from the bulk to the discontinuity surface using the equations for a constant-density fluid. In general, there is no guarantee that such continuation exists, but it must exist for  sufficiently small $l_f.$ Indeed, the field $\bm{v}$ has no singularities in the region where the gas density is constant, that is at distances $\gg l_f$ from the front, as it coincides there with the true velocity field $\bm{V}.$ Therefore, $\bm{v}$ is nonsingular also in a vicinity of this region, and the surface satisfying Eq.~(\ref{traveltimef}) will fall within this vicinity for a sufficiently small $l_f.$ Thus, the field $\bm{v}$ can be continued up to the surface; restriction of $\bm{v}$ to this surface is what was denoted above as $\bm{v}_-.$ In quite the same manner, the field $\bm{v}$ describing the burnt gas velocity can be continued from the bulk region downstream the flame to the surface of discontinuity, possibly at the cost of choosing a still smaller $l_f.$ Finally, it is not difficult to see that within the first order approximation, the functions $\bm{R}(t), \bm{r}(t)$ obtained this way coincide in the bulk downstream of the flame. This is because the tangential velocity components of $\bm{V},\bm{v}$ differ at most by terms of the first order in $l_f,$ and of the same order is the transit time of the gas element through the region of varying density. Neglecting second-order terms, therefore, trajectories $\bm{R}(t),\bm{r}(t)$ run over the same set of points in this region, despite the fact that at any given instant $t,$ the deviation $|\bm{R}(t)-\bm{r}(t)|$ is only of the first order in $l_f.$ But their transit times are also equal, by virtue of Eq.~(\ref{traveltimef}), so that the initial conditions for the two sets of Eq.~(\ref{trajectories}) to be solved in the bulk of burnt gases are identical, as are their right-hand sides. On account of the uniqueness theorem for the normal system of ordinary differential equations, the functions $\bm{R}(t)$ and $\bm{r}(t)$ then coincide in the bulk, thus completing the proof.

To conclude, the similarity between flame fronts and contact surfaces is only superficial. The hydrodynamic problem of flame propagation requires far more detailed description of matter than the thermodynamic problem of equilibrium of heterogeneous substances. The requirement of invariance of the bulk gas motion under the replacement of a real flame with a discontinuity surface uniquely fixes its position. That this replacement is self-consistent at least in the first order in the flame thickness is guaranteed by the proof just given.

\subsection{Vorticity production and velocity jumps at the front}

By virtue of the mass conservation, the normal gas velocity $V^n$ as defined above Eq.~(\ref{traveltimef}) is related to the true gas density, $D ,$ by
$$V^nD  = 1 + O(l_f),$$ where the fresh gas density is used as a density unit. Therefore, Eq.~(\ref{traveltimef}) can be rewritten within its accuracy as
\begin{eqnarray}\label{density}
\int_{-\infty}^{0}d\xi\left(D  - 1\right) + \int_{0}^{+\infty}d\xi\left(D  - \frac{1}{\theta}\right) = 0,
\end{eqnarray}
\noindent where $\xi$ is chosen so that the discontinuity surface is at $\xi = 0,$ $\xi=-\infty$ being in the bulk of fresh gas. The left hand side of this equation is nothing but the so-called density integral.~\cite{class2003} Though the authors of Ref.~\cite{class2003} insist that the discontinuity surface can be arbitrarily placed within the front, they recognized that the vanishing of the density integral entails considerable simplification of the jump conditions at the surface. We can therefore directly use their results obtained under the condition (\ref{density}). First of all, this equation implies that the mass flux, hence, the tangential velocity component is continuous at the surface:
\begin{eqnarray}\label{vtau}
v^{\tau}_- = v^{\tau}_+.
\end{eqnarray}
\noindent Furthermore, the normal component of the burnt gas velocity is simply related to $v^n_-,$
\begin{eqnarray}\label{vnormal}
v^n_+ = \theta v^n_-.
\end{eqnarray}
\noindent Finally, the jump of gas pressure reads
\begin{eqnarray}\label{pressure}
[p] = - v^n_-[v^n] - 2 I_{\sigma}\left(\frac{f'}{N}\right)',
\end{eqnarray}
\noindent where $$I_{\sigma} = \int_{-\infty}^{0}d\xi\left(1 - \frac{1}{D }\right) + \int_{0}^{+\infty}d\xi\left(\theta - \frac{1}{D }\right) = O(l_f)$$ (for brevity, the viscous term is omitted in $I_{\sigma},$ as the finite front-thickness contributions proportional to the front curvature will be found negligible in the subsequent applications anyway). With the help of these relations the vorticity jump can be found using the general formula\cite{hayes}
$$[\sigma] = \frac{1}{Nv^n_-}\left\{\frac{\theta - 1}{\theta}\left(gf' - (v^{\tau})'v^{\tau}\right) - [v^n](v^n_-)' + \left([p] + v^n_-[v^n]\right)'\right\},$$ where $g$ is the gravity acceleration defined positive for flames propagating upwards (this formula is a consequence of Eq.~(\ref{vtau}) and the Euler equations). Since the flow is potential upstream the front, by virtue of the Thomson theorem, this yields the memory kernel $M = Nv^n_+\sigma_+$ in the form
\begin{eqnarray}\label{mkernel}
M = \alpha\left\{gf' - (v^{\tau})'v^{\tau}\right\} - \alpha\theta(v^n_-)' - 2\theta I_{\sigma}\left(\frac{f''}{N^3}\right)' + O(l^2_f), \quad \alpha\equiv \theta - 1.
\end{eqnarray}
\noindent At last, the jump of complex velocity can be written as
\begin{eqnarray}\label{compvel}
[\omega] = \frac{1 - if'}{N}\alpha v^n_-.
\end{eqnarray}
\noindent

\subsection{Assessment of the finite front-thickness effects in vertically propagating flames}\label{assessment}

In order to get insight into the role of the finite front-thickness contributions, it is instructive to look at the equations governing steady propagation of zero-thickness flames driven by strong longitudinal gravity. They can be easily obtained from Eqs.~(3)--(5) of Ref.~\cite{kazakov4}, which describe zero-thickness flames accelerating in a horizontal channel, in the reference frame attached to the front. In view of the equivalence of acceleration and uniform gravity, the substitution $g = 0,$ $a \to g$ ($a$ is the flame acceleration in the laboratory frame) converts them into equations for a flame propagating steadily in a vertical channel, in the laboratory reference frame,
\begin{eqnarray}\label{masteru0}&&
\frac{d}{dx}\left[\frac{u'_-}{f'} (1 + \alpha x) - \frac{\alpha  x g}{u_-}\right] + \alpha\left(\frac{u'_-}{u_-} - \frac{gf'}{u^2_-}\right)(w_- - \alpha) = 0\,, \\&&
f'(x)w_-(x) = (1 - x)u_-'(x)\,,\label{masterw0}\\&&
u_- = f'(w_- + 1)\,.\label{evolutioneq0}
\end{eqnarray}
\noindent This system applies to flames with elongated fronts ($|f'|\gg 1$) driven by strong gravity ($g \gg 1$). It is easily seen that the rescaling $f'\to gf',$ $u_- \to gu_-,$ $w_- \to w_-$ eliminates $g$ from this system. This means that the front slope, hence the flame propagation speed relative to the fresh mixture far upstream, $U,$ scales with $g$ as $U \sim g,$ or, switching for a moment to the ordinary units,
\begin{eqnarray}\label{uscaling}
U \sim \frac{gb}{U_f}\,.
\end{eqnarray}
\noindent Before discussing the scaling of $U$ with respect to $U_f,$ predicted by this formula, it is worthwhile to look at the magnitude of its right-hand side in one particular instance. The speed of near-limit methane-air flames ($U_f = 5-7$\,cm/s) propagating upward in the standard flammability tube of $5.1$\,cm diameter is about $25$\,cm/s, whereas $U$ estimated using Eq.~(\ref{uscaling}) is more than an order of magnitude higher. Yet, it would be hasty to conclude that (\ref{uscaling}) is useless as a flame speed estimate. The point is that solutions of the system (\ref{masteru0})--(\ref{evolutioneq0}) are not unique. Numerical analysis shows that in the case $U_f = 5$\,cm/s and $\theta=5$ this system does have a solution with a very high speed $U=345$\,cm/s, that is of the magnitude suggested by (\ref{uscaling}), but in addition to that, two solutions with $U=25$\,cm/s and $U=29$\,cm/s (identification of the physical solutions is described in Sec.~\ref{identification}). Which of these solutions is realized in practice is a question of their stability and the ignition source strength, as will be discussed later on. Dependence of the flame speed on the fuel concentration is a matter of the same sort. It should be stressed in this connection that the scaling (\ref{uscaling}) holds for a fixed $\theta.$ Therefore, even with the stability issue put aside, relation (\ref{uscaling}) gives little idea about this dependence, because $\theta$ noticeably varies with the mixture composition, whereas the speed of vertically propagating flames is rather sensitive to the density contrast at the front.

Still, the scaling (\ref{uscaling}) holds true for each family of solutions with a fixed $\theta,$ and the form of the dependence $U(U_f)$ raises the question of validity of the zero front-thickness approximation. The point is that according to Eq.~(\ref{uscaling}), the flame propagation speed ought to grow without bound as its normal speed decreases. Therefore, this approximation must eventually break down as $U_f \to 0$ no matter how small the front thickness is. To be more specific, let us use Eq.~(\ref{sfunctional}) to estimate the width $\delta$ of the region near the flame tip where the finite front-thickness effects are significant, that is, $S$ in Eq.~(\ref{evolution}) is of the order of the velocity unit (this region is near the channel centerline when the flame is symmetric, as is usually the case with limit flames; otherwise, it is adjacent to one of the channel walls). To begin with, the term with the front curvature in $S$ can be safely omitted, because flames under consideration are observed to have radii of the front curvature comparable to the tube radius, which in practice is about $10^2$ times larger than the front thickness. In fact, the subsequent numerical analysis confirms that for methane-air flames in the standard flammability tube, for instance, the relative value of the curvature contribution is less than one percent. Next, we note that $\delta$ can be characterized as the distance at which the tangential velocity component grows from zero to a value of the order of $U.$ Assuming that $\delta \ll b$ [hence, the estimate (\ref{uscaling}) holds true], and that $f'$ vanishes at the origin so that the front slope is not large for $x<\delta,$ integration of the relation $S \sim U_f$ over this distance along the front gives $\mathscr{L}_s gb/U_f \sim U_f\delta,$ or
\begin{eqnarray}\label{deltatob}
\frac{\delta}{b} \sim \frac{g\mathscr{L}_s}{U^2_f}\,.
\end{eqnarray}
\noindent It follows, first of all, that the fraction of the front where the finite front-thickness effects are not negligible is independent of the channel width. Second, this fraction itself is not negligible for most hydrocarbon-air flames, except in near-stoichiometric mixtures. Taking again a limit methane-air flame for illustration, substitution of $\mathscr{L}_s \approx - 0.6$\,mm, $U_f\approx 5$\,cm/s in the right-hand side of Eq.~(\ref{deltatob}) gives a value about $2.$ Therefore, the zero front-thickness approximation is entirely inadequate in the case of a fast flame described by the solution with $U=345$\,cm/s. On the other hand, according to the observation following Eq.~(\ref{uscaling}), this relation can serve as a speed estimate for the other two solutions only if a fairly small proportionality coefficient is introduced in its right-hand side. This coefficient would then reappear on the right of (\ref{deltatob}), yielding an order-of-magnitude smaller value of $\delta.$ In other words, in the case of the slow solutions the finite front-thickness effects are significant only in a comparatively small fraction of the front near $x=0.$

We thus arrive at the conclusion that the flame propagation speed in a
vertical tube, regarded as a function of the normal flame speed, is a resultant of two opposite tendencies. On the one hand, decreasing the normal flame speed enhances gravity impact (Froude number increases), which leads to the flame elongation, hence increase of the ratio $U/U_f.$ On the other hand, the growing finite front-thickness effects tend to temper this increase by effectively reducing the portion of the channel cross-section where the front slope is large. The resulting scaling of $U$ with $U_f$ is naturally expected to be modified so as to remove the unbounded growth of $U$ at small $U_f.$ It should be noted, however, that the difficult question of how this modification is effected may well turn out to be practically irrelevant. This is because the growing flow strain will most probably lead to flame quenching, or make the given propagation regime unstable. But even if the question of the flame speed limit at $U_f = 0$ were of practical importance, it could not be settled within the framework of asymptotic expansion in $l_f.$ In fact, this expansion assumes sufficient smallness of the finite front-thickness effects, whereas according to the above estimate, they grow large as $U_f$ decreases. On the purely experimental side, observations\cite{levy1965,vonlavante} indicate that in the limit mixtures, $U \sim \sqrt{gb}.$ This law will be derived numerically in Sec.~\ref{inflammability} as an approximate scaling of the two slower solutions of the master equation, and the issue of their stability will play important role in the theory of partial flame propagation.\cite{footnote1}

\subsubsection{A possibility for dynamical reduction of $\delta$}\label{reduction}

The estimate (\ref{deltatob}) was obtained under the standard assumption that $f' = 0$ at $x=0,b.$ This condition, which ensures vanishing of the transversal velocity component at these points in the burnt gas flow, is essentially kinematic in that it is independent of the combustion regime. At the same time, Eq.~(\ref{compvel}) shows that the vanishing of $w,$ say at $x=0,$ can be achieved also for $f' \ne 0,$ provided that the combustion regime is such that the burning rate vanishes at this point. Indeed, in this case $v^n_- = 0,$ and hence $$[w] = - \frac{\alpha f'}{N}v^n_- = 0,$$ so that $w_+$ turns into zero together with $w_-.$ In other words, the gas velocity gradients generated by the strong flame-gravity interaction near $x=0$ are large enough to terminate the reaction. Of course, existence of such conditions is a non-perturbative matter which cannot be settled within the small-$l_f$ expansion. Realization of these conditions would make possible a situation where $\delta$ estimated according to (\ref{deltatob}) is comparable to the channel width, but the finite front-thickness effects are suppressed everywhere except in a small vicinity of $x=0$ nevertheless. Indeed, if the front slope is allowed to be large already at $x=0,$ then the tangential velocity component could reach $O(U)$ values at arbitrarily small distances from $x=0,$ whereas the condition $v^n_- = 0$ at $x=0$ would guarantee consistency of the on-shell equations everywhere in the channel including the origin. Such regimes then would be well-described by the zero front-thickness Eqs.~(\ref{masteru0})--(\ref{evolutioneq0}) no matter how large $U$ is.

It is to be noted that for flames symmetric with respect to $x=0,$ the vanishing of $f'(0)$ is a consequence of the flame symmetry, so that the described possibility pertains only to asymmetric flame patters where the line $x=0$ is the channel wall.

\section{Equations for finite thickness flames}\label{mainequations}

\subsection{Large-slope expansion of the master equation}\label{largeslope}

Observations show that in sufficiently wide tubes, the flame propagation speed largely exceeds the normal flame speed, which means that the flame front slope is large for the most part of the tube cross-section, $|f'|\gg 1.$ As discussed in Sec.~\ref{assessment}, equations describing vertical flame propagation in this case are expected to have solutions with significantly different propagation speeds. In particular, the flame speed can be so large that the finite front-thickness effects are strong all along the front. In these circumstances, expressions (\ref{mkernel}), (\ref{compvel}) do not admit any sensible simplification, and are to be substituted into Eqs.~(\ref{master1}), (\ref{evolution}) as they stand, which yields very cumbersome equations. It is doubtful that these equations can be of practical significance, because the small-$l_f$ expansion is invalid in a situation where $\delta\gtrsim b$ anyway. At the same time, the finite front-thickness effects are moderate on the slower solutions even near the limits of inflammability, being confined in a comparatively small fraction of the front near the flame tip. This non-uniformity opens a simple way of taking these effects into account. Namely, one has to first identify contributions which do not involve factors of $f',$ $w,$ {\it etc.} (that is, odd functions of $x$). It is these contributions that are of potential to significantly change the flame structure near its tip; in this region, $f',$ $w$ are small, and therefore the relevant terms can be used in a linearized form. On the other hand, the comparatively small effect of contributions proportional to $f',$ $w$ in the region where $|f'|\gg 1$ can be incorporated by replacing them with the leading terms of the large-slope expansion. The subsequent analysis follows closely that given in Ref.~\cite{kazakov3} for horizontal flame propagation. Consider the region where the front slope is large, $|f'|\gg 1.$ Under this condition, the $\hat{\EuScript{H}\,}$-operator appearing in Eq.~(\ref{master1}) simplifies to
\begin{eqnarray}\label{hcurvedf1}
\left(\hat{\EuScript{H}\,}a\right)(x) = (f'(x) -
i)~\int_{0}^{1} d\eta~\frac{a(\eta) +
a(-\eta)}{2}\chi(\eta - |x|)+ia(-x)(2|x| - 1) +
O\left(\frac{1}{f'}\right),\nonumber
\end{eqnarray}
\noindent where $\chi(x)$ is the sign function,
$$\chi(x) = \left\{
\begin{array}{cc}
+1,& x>0\,,\\
-1,&  x<0\,.
\end{array}
\right.
$$ This expansion is sufficient to extract the leading term of Eq.~(\ref{master1}), which is contained in its real part. The only contribution proportional to $l_f$ that requires special care is that coming from the jump of $w$ velocity component. Using Eq.~(\ref{compvel}) with $v^n_-$ from Eqs.~(\ref{evolution}), (\ref{sfunctional}), the leading term of $\hat{\EuScript{H}\,}[w]'$ is readily found to be proportional to the integral
$$\int_{0}^{1}d\eta \left(\frac{f'}{N}\frac{\mathscr{L}_s v'_{\tau}}{N}\right)'(\eta)\chi(\eta - |x|) = \left.\frac{\mathscr{L}_s f'v'_{\tau}}{N^2}\chi(\eta - |x|)\right|_{0}^{1} - \frac{2\mathscr{L}_s f'v'_{\tau}}{N^2}(|x|).$$ The first term on the right vanishes at $\eta = 0,$ because $f'$ is exactly zero there, whereas $$\left.\frac{f'v'_{\tau}}{N^2}\right|_{\eta=1} \approx \left.\frac{du_-}{df}\right|_{\eta = 1} \equiv u'_1.$$ Since the last term is proportional to $f',$ it is to be replaced by its leading-order expression $$\frac{\mathscr{L}_sf'v'_{\tau}}{N^2} \to \frac{\mathscr{L}_s u'_-}{f'}.$$ Extraction of the leading term of Eq.~(\ref{master1}) thus yields
\begin{eqnarray}
2(1 + \alpha \mathscr{L}_s)u'_- + 2\alpha |x|\left(u_-' - \frac{gf'}{u_-}\right) - \alpha f'(1 + \mathscr{L}_s u'_1) + f'\int_{0}^{1}d\eta \frac{Mw_+}{v^2_+}[\chi(\eta - |x|) - 1] = 0.\nonumber\\ \label{master2}
\end{eqnarray}
\noindent The integrand in the last term is not expanded because the integration is over the whole channel cross-section, including the regions where the front slope is not large. To get rid of this integral, we divide Eq.~(\ref{master2}) by $f',$ and then differentiate it with respect to $x$
\begin{eqnarray}
\frac{d}{dx}\left[(1 + \alpha \mathscr{L}_s)\frac{u_-'}{f'} + \alpha |x|\left(\frac{u_-'}{f'} - \frac{g}{u_-}\right)\right] - \frac{Mw_+}{v^2_+}\chi(x) = 0.\nonumber
\end{eqnarray}
\noindent The last term is now a local function, so that it can be expanded as the rest of the equation. It is convenient to rewrite the result taking $f$ as an independent variable, and to limit consideration by positive $x$'s
\begin{eqnarray}
\frac{d}{df}\left[(1 + \alpha \mathscr{L}_s)\frac{du_-}{df} + \alpha x(f)\left(\frac{du_-}{df} - \frac{g}{u_-}\right)\right] + \frac{\alpha}{u_-}\left(\frac{du_-}{df} - \frac{g}{u_-}\right)\left(w_- - \alpha + \alpha\mathscr{L}_s\frac{du_-}{df}\right) = 0. \nonumber\\\label{master3}
\end{eqnarray}
\noindent The other equation can be obtained from a dispersion relation for the upstream velocity
\begin{eqnarray}\label{chup}
\left(1 - i \hat{\EuScript{H}\,}\right)\omega_-' = 0\,,
\end{eqnarray}
\noindent which is also a consequence of Eq.~(\ref{master1}), as is seen by applying $(1 - i \hat{\EuScript{H}\,})$ to its left-hand side and using the exact identity $\hat{\EuScript{H}\,}^2 = - 1.$ As no flame specifics is involved in Eq.~(\ref{chup}), its expansion is still given by Eq.~(\ref{masterw0}), or
\begin{eqnarray}\label{3rel}
w_- = (1 - x)\frac{du_-}{df}\,.
\end{eqnarray}
\noindent Equations (\ref{master3}), (\ref{3rel}) are ordinary differential equations which can be combined to give a second order differential equation for the function $u_-(f),$ provided that the function $x(f)$ is given. It is easily seen that as long as $x(f)$ is continuous and $u_-$ is positive, $d^2u_-/df^2$ defined by this equation is a continuous function of $f$ and $du_-/df,$ if $(1 + \alpha \mathscr{L}_s + \alpha x)$ is positive on the interval $x\in (0,1).$ Since $x$ is positive, this amounts to the requirement
\begin{eqnarray}\label{positivity1}
1 + \alpha \mathscr{L}_s >0,
\end{eqnarray}
\noindent which is practically always satisfied despite $\mathscr{L}_s$ can be negative and $\alpha$ can be large ($\alpha$ normally takes on the values 4--7 for hydrocarbon-air flames, but can be larger in undiluted mixtures). Therefore, under the specified conditions on $x(f)$ and $u_-,$ a unique solution to the second-order differential equation exists, provided that initial conditions for $u_-$ and its derivative are specified at some point $x_0.$ As in the horizontal case, this fact can be used to continue the solution over the whole interval $x \in (0,1),$ to avoid the necessity of separate consideration of the region near $x = 0$ where Eqs.~(\ref{master3}), (\ref{3rel}) are not valid because of the smallness of the front slope. To this end, we note that any solution to Eqs.~(\ref{master2}), (\ref{3rel}) satisfies also Eq.~(\ref{master3}), therefore, it is unique under the initial conditions it defines, say, at the point $x_0 = 1/2.$ Extension of Eq.~(\ref{master2}) over $x\in (0,1)$ then uniquely extends the given solution over this domain. Equation~(\ref{master2}) can also be used to replace the initial conditions for a continued solution, existing at $x_0$ inside the channel, by boundary conditions at its walls. Namely, one such condition, which merely accounts for the rise of a differential order in the transition from Eq.~(\ref{master2}) to Eq.~(\ref{master3}), can be obtained by setting $x=0$ in the former equation. The point is that the last (integral) term in Eq.~(\ref{master2}), which depends on the unknown flow structure near $\eta=0,L,$ vanishes at $x=0.$ Denoting $du_-(0)/df\equiv u'_0,$ one finds
\begin{eqnarray}\label{boundaryc1}
u'_0 = \frac{\alpha(1 + u'_1\mathscr{L}_s)}{2(1 + \alpha \mathscr{L}_s)}\,.
\end{eqnarray}
\noindent Notably, derivatives of the gas velocity at the two channel walls turn out to be interrelated, which is a reflection of the inherent nonlocality of the combustion process. Equation~(\ref{boundaryc1}) is thus really not an initial, but a boundary condition. It is to be stressed, however, that the coupling between the boundary values of $du_-/df$ must be weak, which is part of the general requirement on the finite front-thickness corrections be sufficiently small within the asymptotic expansion in $l_f.$

The other boundary condition valid at the leading order of the large-slope expansion is \cite{kazakov3}
\begin{eqnarray}\label{boundaryc2}
u_-(0) = U.
\end{eqnarray}
\noindent
For brevity, we do not introduce special designation for the functions continued in the way just described; this should not lead to confusion as their originals (that is, functions before continuation) will appear again only in the course of transformation of the evolution equation in the rest of the present section, where they will be explicitly identified as such.

The constructed continuation was seen to be unique for each given continuous function $x(f).$ As this function itself is a functional of the on-shell velocity, whose structure is essentially determined by the evolution equation, it remains to be shown that this functional does not subvert consistency of the system (\ref{master3})--(\ref{boundaryc2}). With $S$ given by Eq.~(\ref{sfunctional}), the evolution equation (\ref{evolution}) reads, in terms of the flow variables before continuation,
\begin{eqnarray}\label{evolutioneq1}
u_- - f' w_- = N - \mathscr{L}_s\left(\frac{w_- + f'u_-}{N}\right)'.
\end{eqnarray}
\noindent It is easy to see that in the term $(w_-/N)'$ on the right, only $w_-$ is to be differentiated in effect. Indeed, the tangential velocity gradient near the flame tip is large, whereas $w_-$ itself is small, so that one can replace $(w_-/N)' \to w_-'$ in this region. On the other hand, both $(w_-/N)'$ and $w_-'$ are small compared to $N$ in the region where $|f'|$ is large. Therefore, taking into account the small factor $\mathscr{L}_s,$ this replacement makes a fairly negligible change of the right-hand side of Eq.~(\ref{evolutioneq1}) wherever the front slope is large. As to the last term in Eq.~(\ref{evolutioneq1}), it can be safely omitted because it is uniformly small, and so in no way modifies the structure of the evolution equation. In fact, writing $$\mathscr{L}_s\left(\frac{f'u_-}{N}\right)' = \mathscr{L}_su_-\left(\frac{f'}{N}\right)' + \mathscr{L}_s u_-'\frac{f'}{N}\,,$$ one sees that the first term represents a relative correction $\leqslant \mathscr{L}_s$ to the coefficient of $u_-$ on the left of Eq.~(\ref{evolutioneq1}); likewise, the term $\mathscr{L}_s u_-'f'/N$ could be noticeable only away from the flame tip, where it is $\approx\mathscr{L}_s u_-',$ but using Eq.~(\ref{3rel}) to write $f'w_- = (1-x)u'_-$ for the second term on the left of Eq.~(\ref{evolutioneq1}) readily shows that this contribution is equivalent to a change of the channel width from $1$ to $(1 - \mathscr{L}_s).$ Therefore, retaining this term would not make much sense, as the effect of such a change is hardly detectable experimentally. Evolution equation thus takes the form
\begin{eqnarray}\label{evolutioneq2}
u_- - f' w_- = N - \mathscr{L}_s w'_-.
\end{eqnarray}
\noindent We now have to rewrite this equation in terms of the functions continued with the help of Eqs.~(\ref{master3}), (\ref{3rel}). As before, one starts from the region with $|f'|\gg 1$ where the two sets of variables coincide. Using Eq.~(\ref{3rel}) to eliminate $w_-$ from Eq.~(\ref{evolutioneq2}), replacing $N\to f'$ on the right of this equation, and then integrating gives
$$- u_-(1 - x) = f - \mathscr{L}_s(1 - x)\frac{du_-}{df} + {\rm const}.$$ With the understanding that the flow variables in this equation are continued to $x=0,$ the constant of integration is fixed by the boundary conditions (\ref{boundaryc1}), (\ref{boundaryc2}) for the continued functions. Taking into account also that $f=0$ at $x=0,$ we find
$${\rm const} = \frac{\alpha\mathscr{L}_s}{2(1 + \alpha\mathscr{L}_s)} - U,$$
and finally arrive at the following expression for $x(f)$ as a functional of $u_-$
\begin{eqnarray}\label{x(f)}
x(f) = 1 - \frac{U - f - \alpha\mathscr{L}_s/2(1 + \alpha\mathscr{L}_s)}{u_- - \mathscr{L}_s du_-/df}\,.
\end{eqnarray}
\noindent It is important that the right-hand side involves only $u_-$ and $du_-/df,$ but not higher derivatives of $u_-.$ This guarantees that upon substitution of the expressions (\ref{x(f)}), (\ref{3rel}) into Eq.~(\ref{master3}), the resulting equation for $u_f$ is still a second-order differential equation, so that conditions (\ref{boundaryc1}), (\ref{boundaryc2}) are sufficient to fix its solution. It is to be noted also that in addition to (\ref{positivity1}), the existence of a solution is further conditioned by the requirement that
\begin{eqnarray}\label{positivity2}
0 \leqslant \frac{U - f - \alpha\mathscr{L}_s/2(1 + \alpha\mathscr{L}_s)}{u_- - \mathscr{L}_s du_-/df} \leqslant 1,
\end{eqnarray}
\noindent following from the fact that $x\in [0,1].$ In particular, $(u_- - \mathscr{L}_s du_-/df)$ must not vanish on this segment, which can be violated in too narrow channels or too strong flame stretch. With these conditions satisfied, this proves consistency of the above system of equations.

It will be noted that the functions we work with from now on are obtained by means of two different continuation procedures. The first is a continuation of the true velocity field $\bm{V}$ from the bulk to the surface of discontinuity defined by Eq.~(\ref{traveltimef}) using the equations for a constant-density fluid, and the second, the continuation just described of the on-shell value of this field and of the function describing position of the discontinuity surface from inside the channel over the near-wall region.

\subsection{Identification of physical solutions}\label{identification}

As is well-known, equations for the flame front position, obtained by integrating the fundamental flow equations, require a supplementary condition to fix an integration constant. Such are the nonlinear Sivashinsky-Clavin equation for weakly curved flames\citep{sivclav} and its higher-order corrections,\cite{kazakov2,kazakov5} Markstein equation for a flame in a time-dependent gravitational field,\cite{markstein1964} {\it etc.} In all these equations, the integration constant is to be found by averaging the equation across the channel. The master equation requires a similar condition, but it cannot be imposed directly on Eq.~(\ref{master1}) before it is integrated. It was conjectured in Ref.~\cite{kazakov3} that this condition can be formulated as the requirement that the average of
$$\frac{M(\eta)\omega_+(\eta)}{v^2_+(\eta)}$$ across the channel vanish.
Using formal expansion with respect to $\alpha,$ it can be shown that this average is zero indeed to at least the fourth post-Sivashinsky approximation (which corresponds to retaining terms of the seventh order in $\alpha$ in the master equation), but a general proof is still lacking. However, the use of this condition for the functions continued as described in Sec.~\ref{largeslope} is problematic anyway, because an expression to be averaged involves higher spatial derivatives of the flow variables, so that the small region near $x=0$ where the gas velocity gradients are large gives rise to a non-small contribution to the average. To overcome this difficulty, one can use a condition of the total flow momentum (or energy) conservation through the transition domain, $y \in (0,U),$ which is satisfied identically by the original solutions since the master equation is an exact consequence of the fundamental flow equations. This condition is free of the higher-derivative terms, so that it can be imposed on the continued functions as well.

Thus, physical solutions will be identified as those conserving the total flow momentum parallel to the walls. An analytical expression of this condition can be obtained as follows. Introducing the reduced pressure $\tilde{p} = p + \phi$ of the fresh gas, and $\tilde{p} = p + \phi/\theta,$ that of the combustion products, where $p$ is the gas pressure and $\phi = - gy$ the gravitational potential, conservation of the longitudinal momentum component in the upstream and downstream parts of the transition domain gives, respectively,
\begin{eqnarray}\label{momentumup}
\tilde{p}|_{y=0} + U^2 &=& \int_0^1d\eta[\tilde{p}_-(\eta) + Nv^n_-u_-(\eta)]\,, \\
\int_0^1d\eta[\tilde{p}_+(\eta) + \frac{N}{\theta}v^n_+u_+(\eta)] &=& \tilde{p}|_{y=U} + \frac{1}{\theta}\int_0^1dx u^2(x,U)\,, \label{momentumdown}
\end{eqnarray}
\noindent The reduced pressure can be eliminated with the help of Bernoulli integrals for the fresh and burnt gas flows (neglecting $w$ in comparison with $u$ in the leading order of the large-slope expansion)
\begin{eqnarray}
\tilde{p}_-(\eta) + \frac{u^2_-(\eta)}{2} &=& \tilde{p}|_{y=0} + \frac{U^2}{2}\,, \label{bernoulliup}\\
\theta\tilde{p}_+(\eta) + \frac{u^2_+(\eta)}{2} &=& \theta\tilde{p}|_{y=U} + \frac{u^2(x(\eta),U)}{2}\,,
\label{bernoullidown}
\end{eqnarray}
\noindent where $(x(\eta),U)$ is the intersection of the line $y=U$ with the streamline that crosses the front at $(\eta,f(\eta)),$ Fig.~\ref{fig1}. Finally, recalling Eqs.~(\ref{vnormal}), (\ref{pressure}) [neglecting, as before, the finite front-thickness correction due to the front curvature in the pressure jump], and using the burnt gas flow continuity
\begin{eqnarray}
v^n_+ N(\eta)d\eta &=& u(x(\eta),U)dx,\label{continuity}
\end{eqnarray}
\noindent Eqs.~(\ref{momentumup}), (\ref{momentumdown}) can be combined into one
\begin{eqnarray}\label{phys}
\int_0^1d\eta N v^n_-u(x(\eta),U) = \frac{u^2_1 + U^2}{2} + \frac{\alpha g}{\theta}\int_0^1d\eta[f(\eta) - U] + \alpha.
\end{eqnarray}
\noindent Here $u_1$ is the value of the longitudinal component of the fresh gas velocity at the front endpoint $(1,U),$ and
\begin{eqnarray}\label{uvelocity}
u(x(\eta),U) = \left\{\theta u^2_1 - \alpha u^2_-(\eta) + 2\alpha g [f(\eta) - U] + 2\alpha\theta \left[1 - (v^n_-(\eta))^2\right]\right\}^{1/2}
\end{eqnarray}
\noindent the burnt gas velocity at the lower boundary of the transition domain. Equation (\ref{phys}) is the sought condition for selecting physical solutions.

\subsection{Numerical solutions}

\begin{figure}
\centering
\includegraphics[width=0.45\textwidth]{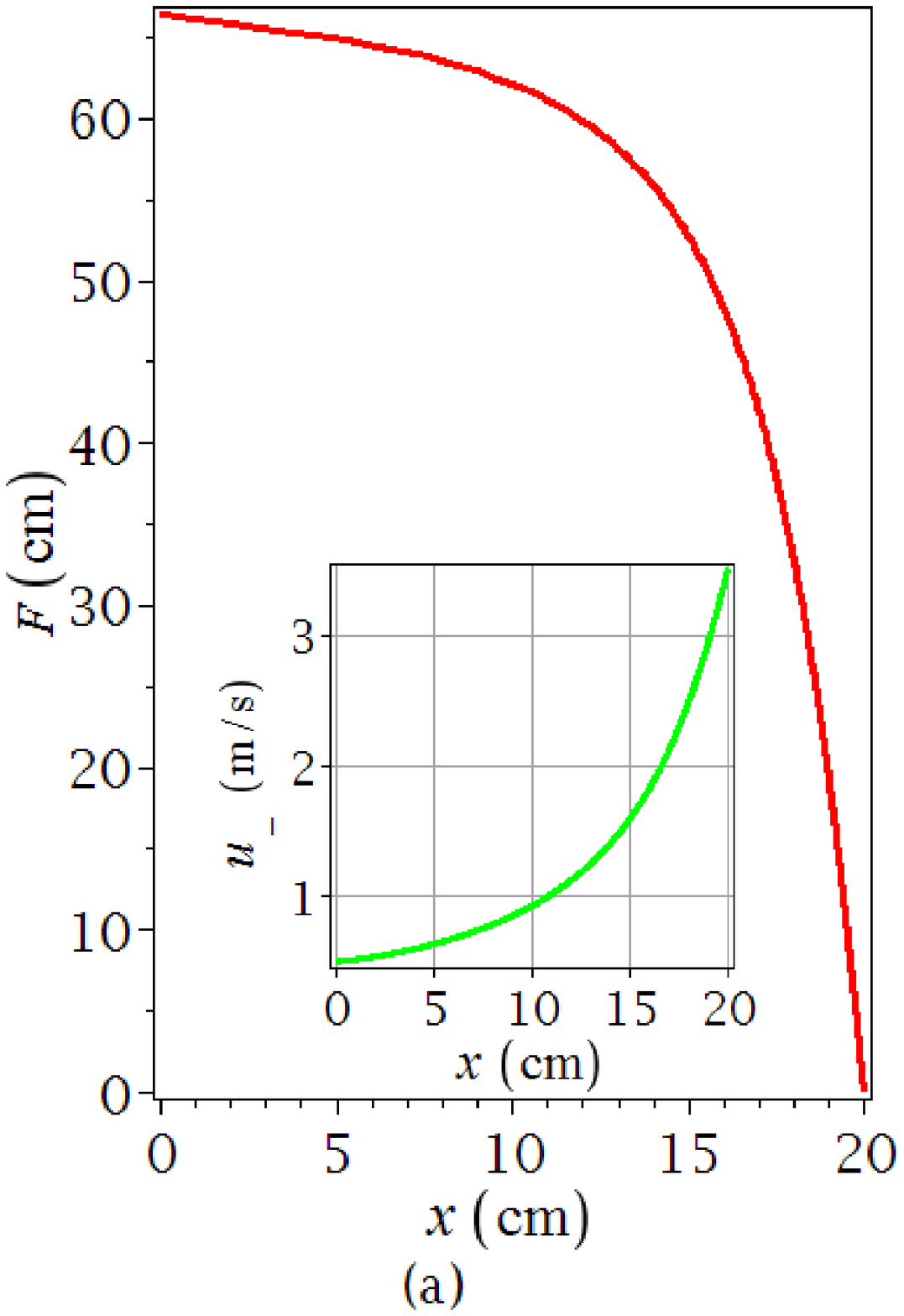}\vspace{2cm}
\includegraphics[width=0.45\textwidth]{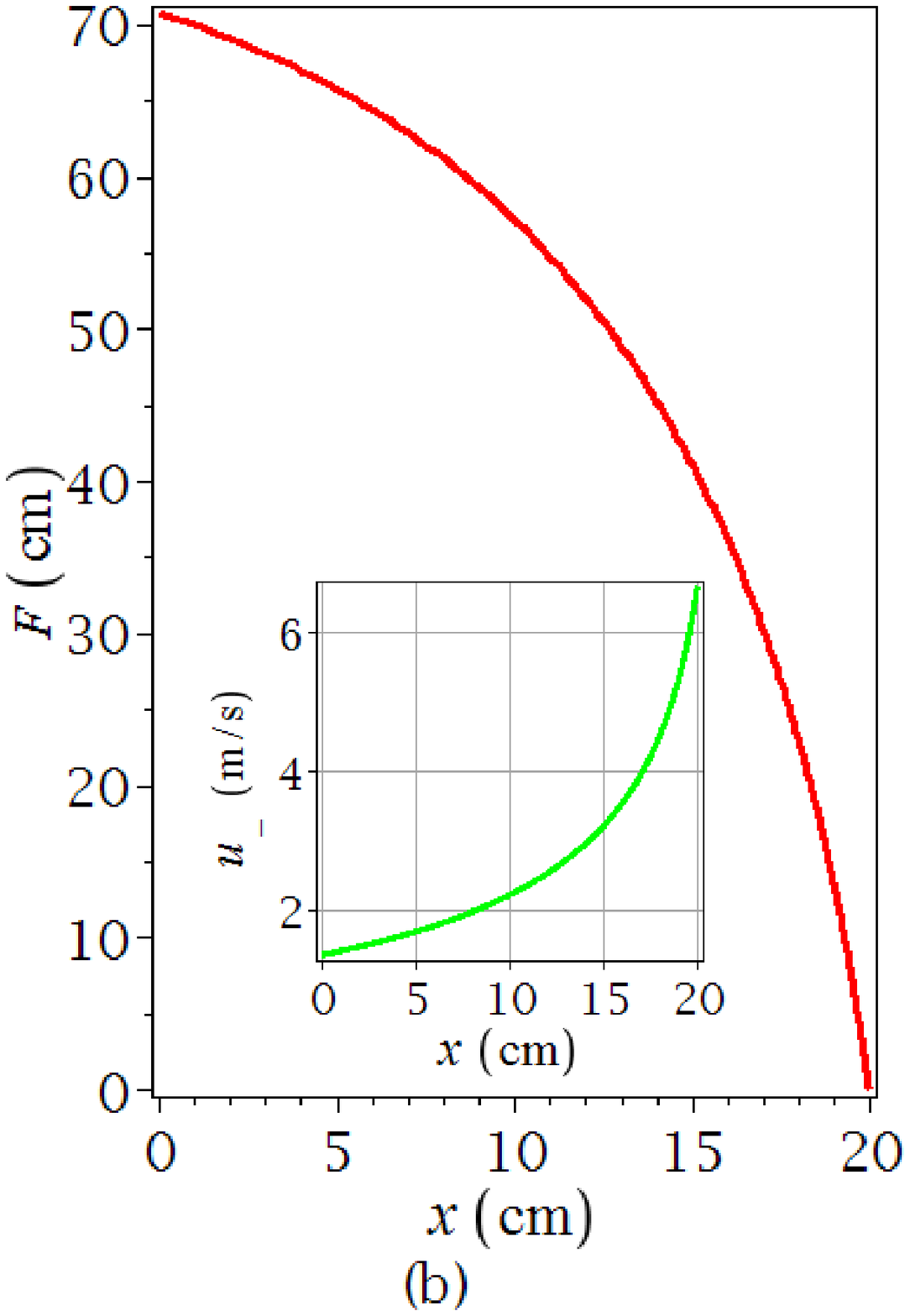}
\caption{(Color online) Front positions $F(x)=f(b) - f(x)$ and on-shell gas velocities $u_-(x)$ (insets) for flames with $\theta = 8,$ $\mathscr{L}_s/b=-0.01$ in a $20$\,cm-wide channel. (a) type I flame with $U_f=15$\,cm/s, (b) type II flame with $U_f=40$\,cm/s.}\label{fig2}
\end{figure}

Numerical scrutiny of the system (\ref{master3}), (\ref{3rel}), (\ref{boundaryc1}), (\ref{boundaryc2}), (\ref{x(f)}) shows that in general, there are two families of solutions with $U$ belonging to two disjoint continua, as in the case of horizontal flame propagation. Unlike the latter, however, the gap between the two sets of $U$'s is much wider. Following the terminology introduced in the horizontal case, solutions having lower (higher) $U$ will be called type I (type II) flames. Among solutions of each type, there can be several physical, that is satisfying condition (\ref{uvelocity}), or there can be none. Front position and longitudinal velocity of physical type I and type II solutions are plotted in Fig.~\ref{fig2}. Recall that these solutions represent functions continued from the large-slope region as described in Sec.~\ref{largeslope}, and so both types of flames have finite front slope at $x=0.$ In other words, it is understood that there is a small region near $x=0,$ where $f'$ and $w$ rapidly drop to zero. Type II flames commonly have significantly larger values $|f'(0)|$ than type I, hence larger velocity gradients at the origin. By this reason, type II regimes is the first instance where one can expect realization of the possibility discussed in Sec.~\ref{reduction}. Also, type I flames are characterized by an inflection of the front near $x=0.2b,$ which is barely seen in Fig.~\ref{fig2}(a), but can be pronounced for other values of the parameters $\theta,U_f,\mathscr{L}_s.$ It is suppressed in the most common case of methane-air mixtures, or other mixtures characterized by $\mathscr{L}_s<0.$ This front inflection is similar to that in horizontally propagating flames, and is a property of the flame two-dimensionality. As the experiment shows, flames propagating in horizontal tubes do tend to flatten, so that steady flames often turn out to be two-dimensional to a high degree, and flames with inflected fronts have been directly observed.\cite{elrabii2015} In contrast, vertically propagating flames are usually observed axisymmetric, or of more complex geometries far from planar, and this raises the question of practical applicability of the two-dimensional analysis to these flames. It is quite clear that such qualitative features as, say, the mechanism of flame extinction by a gravitational field are common for two- and three-dimensional flames. On the quantitative side, to get insight into  how the flame dimensionality affects its speed, it is instructive to compare flame propagation with the propagation of bubbles in a channel and in a tube of equal size. Assuming an axisymmetric geometry, consider first bubble motion, and let a three-dimensional bubble be obtained from the respective two-dimensional channel pattern by rotating it around the tube axis. The bubble speed depends essentially on the relative cross-sectional area of the jet formed between the bubble and the walls, which is evidently larger in the three-dimensional case. The same geometrical reason suggests that if it were possible to obtain an axisymmetric flame by rotating a two-dimensional pattern around the tube axis, then the speed of such a flame would be larger than the speed of its two-dimensional original, because the front slope grows away from the tube axis. However, this rotation would violate the flow continuity, because the gas elements burnt near the tube wall come from regions of lesser radial distance from the axis, whereas the flow far upstream remains homogeneous. Therefore, this flame will reduce its longitudinal spread in order to maintain the flow continuity, rather than increase its propagation speed. On the other hand, retaining the continuity in the case of bubbles of a fixed shape is not a problem, for their motion is not bound to satisfy such condition as a constancy of the local burning rate. We thus see that the difference in the mechanisms governing propagation of bubbles and flames implies their essentially different sensitivity to the flow dimensionality.

Admittedly, these arguments do not affirm equality of the propagation speeds of two- and three-dimensional flames, but they point toward certain rigidity of the flame speed with respect to manifestations of the three-dimensional nature of the problem. On this ground, the speed of a flame propagating in a tube of diameter $d$ will be identified henceforth as the speed of a two-dimensional flame in a channel of width equal to the tube diameter, $b=d.$ Subsequent applications fully confirm this presumption.

\subsection{Viscous energy dissipation and flame stability}\label{dissipation}

It was already mentioned that in general, there are several solutions of the system (\ref{master3}), (\ref{3rel}), (\ref{x(f)}) satisfying condition (\ref{phys}). In other words, there are different regimes of steady flame propagation, characterized by different propagation speeds. This fact raises an important question of their actual realizability, which in turn calls for a stability analysis of the steady solutions. Although this is generally a very difficult task, there is one important case where {\it instability} of a given regime can be deduced directly from the properties of the burnt gas flow generated by a steady flame. Consider the flow vorticity distribution downstream of the flame. Vorticity produced in the flame front gradually decays as the burnt gas moves away from the front. This is a result of two factors -- an internal viscous dissipation in the bulk of the burnt gas flow, and the viscous drag exerted on the flow by the channel walls. The latter takes place in a boundary layer that forms behind the front and grows downstream. Even if the flow in this layer is turbulent, it fills the channel cross-section only at distances largely exceeding the flame size. Therefore, at distances comparable to the flame size, which are only important for the formation of the flame structure, the wall influence is negligible, and the vorticity decays only under the action of internal friction. Let us determine the total rate of kinetic energy dissipation assuming that the vorticity decays completely, so that the burnt gas flow ultimately becomes homogeneous. The energy conservation downstream the transition domain reads
\begin{eqnarray}
q_K|_{y=H} + \theta U\tilde{p}|_{y=H} = q_K|_{y=U} + \theta U\tilde{p}|_{y=U} -\dot{Q}\,,
\label{energydiss1}
\end{eqnarray}
\noindent where $$q_K = \int_0^1 dx u\frac{v^2}{2\theta}$$ is the kinetic energy flux through a horizontal cross-section of the channel, $y=H$ symbolizes sufficiently large $y$'s where the gas flow is nearly homogeneous, and $\dot{Q}$ is the total rate of viscous dissipation of the kinetic energy due to the flow inhomogeneity. Using the conservation of $y$-component of the flow momentum in this domain
\begin{eqnarray}\label{momentuminf}
\tilde{p}|_{y=U} + \frac{1}{\theta}\int_0^1dx u^2(x,U) = \tilde{p}|_{y=H} + \theta U^2
\end{eqnarray}
\noindent to eliminate the reduced pressure from Eq.~(\ref{energydiss1}), and then Eq.~(\ref{continuity}) to express the remaining terms as integrals over the flame front yields
\begin{eqnarray}
\dot{Q} = \frac{\theta^2 U^3}{2} + \int_0^1d\eta Nv^n_-\left[\frac{u^2(x(\eta),U)}{2} - \theta U u(x(\eta),U)\right].
\label{energydiss2}
\end{eqnarray}
\noindent Notice that the right-hand side of this equation is not positive definite, that is, it can be negative on some functions $u(x(\eta),U) \equiv u(\eta).$ Since $\dot{Q}$ must be non-negative, this would mean that the given flow actually does not become homogeneous far downstream. In other words, the bulk viscous forces alone are incapable of transforming the burnt gas flow with this profile into the state of thermodynamic equilibrium -- a homogeneous flow. Thus, the combustion regime producing such flow turns out to be {\it metastable}: the flame generating $u(\eta)$ with $\dot{Q}<0$ acquires a burnt gas ``tail'' in a state of incomplete equilibrium, and sooner or later will undergo transition to a regime with $\dot{Q}>0.$ It will be shown in Sec.~\ref{inflammability} that under certain conditions, physical solutions characterized by $\dot{Q}<0$ do exist and play important role in explaining the observed flame behavior.

\section{Applications}\label{applications}

\subsection{Theory of inflammability\cite{kazakov6}}\label{inflammability}

\subsubsection{Introduction}

A standard tube used to measure inflammability limits is $5.1$\,cm in diameter and $1.8$\,m long. If a mixture ignited at its bottom propagates all the way up the tube, it is said to be inflammable.\cite{cowardjones1952,zabetakis} In view of the practical significance of methane-air and propane-air flames, their limit properties have been thoroughly investigated in tubes of circular and square cross-sections with $b = 5.1,$ 9.5, 10 and  20\,cm.\cite{cowardjones1952,zabetakis,levy1965,vonlavante,jarosinski} The studied flames were characterized by different fresh-to-burnt gas density ratios $\theta = 4.7$ to $\theta = 5.3,$ had different burning rates and molar-mass relations between fuel and oxidizer. Despite these distinctions, a great deal of similarity in the limit flame behavior has been established which can be summarized as follows:

\begin{description}
\item[i] Before extinction, flames can propagate steadily over distances largely exceeding the flame size, at least in tubes with diameter up to $10$\,cm. The onset of extinction is simultaneous with a slight increase in the flame speed.

\item[ii] The inflammability range narrows in wider tubes, that is, the minimal burning rate required to propagate the flame increases with the tube diameter.

\item[iii] The propagation speed of limit flames in a tube of given diameter coincides, within the experimental error, with the speed of an air bubble rising in the same tube filled with water.

\item[iv] Reactants flow into the rising flame, but only around the hot post-flame structure that continues to rise with the same speed after the flame extinction. During extinction, the flame centre vanishes first, followed rapidly by the edges.
\end{description}

Taken together, these observations make the nature of partial flame propagation and its extinction quite a riddle. On the one hand, the apparent flame steadiness before extinction means that the characteristic time of partial flame propagation is much larger than the transit time of gas through the region occupied by the flame (transition domain in what follows). On the other hand, the final stage of the process is much more rapid, according to (iv). The latter would take place if the extinction process were buoyancy driven, as is also suggested by observation (iii), but in that case the flame ought to extinguish right upon entering the steady regime, that is a few tube diameters above the ignition point, in contradiction to the first part of (i). The second observation of (i) is confusing on its own, as the flame which is about to extinguish would rather be expected to decelerate. Next, heat losses to the walls, which are generally an important factor of flame extinction, cannot drive the process under consideration because their relative value diminishes as the tube diameter increases, and so they would produce a trend opposite to (ii). In fact, it is known that the heat losses are negligible for axisymmetric flames propagating in tubes with diameter $\gtrsim 2$\,cm. Neither can extinction be effected by the flame stretch.\cite{vonlavante,buckmaster} In the case of methane-air flames, for instance, the flame stretch increases the burning rate at the flame centre\cite{bradley}, thus leading to contradiction with (iv). This mechanism is also unable to explain (ii), because it would produce opposite trends in mixtures with light and heavy deficient component (such as lean methane-air and propane-air mixtures, respectively), contrary to what is observed.

The mechanism of partial flame propagation and its extinction will be identified below by solving numerically the system (\ref{master3}), (\ref{3rel}), (\ref{x(f)}) subject to the conditions (\ref{boundaryc1}), (\ref{boundaryc2}), (\ref{phys}). The intricacy of properties (i)-(iv) will be shown to be due to a rather nontrivial structure of the spectrum of flame propagation regimes under strong gravity.

\subsubsection{A remark on the value of Markstein length}\label{remark}

As defined in Sec.~\ref{discontinuity}, the functions $u_-(x),w_-(x)$ denoted by lowercase letters describe the gas velocity in the picture where the finite-thickness flame front is replaced by a discontinuity surface. These functions are restrictions to the surface fixed by Eq.~(\ref{traveltimef}) of an auxiliary field $\bm{v}(x,y)$ which is to be found as a continuation of the physical velocity field $\bm{V}(x,y)$ from the bulk using the flow equations for a constant-density fluid. To the best of my knowledge, however, the difference between $\bm{v}$ and $\bm{V}$ is ignored in practical determinations of the Markstein length, which are rather focused on the question where the discontinuity surface is to be placed in order to obtain a best fit of the experimental data.\cite{taylor,davis2002,matalon} Since displacing this surface at a $O(l_f)$-distance within the front gives rise to a $O(l_f)$ change of the Markstein length, and so does neglecting the difference $(\bm{v}-\bm{V}),$ the values obtained this way can serve only as a rough estimate of the parameter $\mathscr{L}_s$ appearing in the equations derived in Sec.~\ref{mainequations}. Furthermore, this parameter is not known but for methane-air flames, as the experiments commonly determine only a single Markstein length, without discriminating the curvature and flow strain contributions. In these circumstances, our primary goal below will be to identify qualitative changes in the flame dynamics brought about by the finite front-thickness effects. For want of more accurate evaluations, the results of Ref.~\cite{bradley}, the only experimental work that distinguishes $\mathscr{L}_s$ and $\mathscr{L}_c,$ will be used in the case of methane-air flames despite the fact that the front is identified in this work not according to Eq.~(\ref{traveltimef}), but as the isotherm $5$\,K above room temperature (it is to be noted also that $\mathscr{L}_s$ is defined in Ref.~\cite{bradley} as the (negative) coefficient of the normal flow strain contribution to the normal flame speed, but since for flames under consideration the curvature effect is negligible, this definition coincides with ours). Together with the figures for propane-air flames, obtained for a pair of reasonably small values of the Markstein length, this will allow determination of the initial trends in the flame behavior with respect to the parameter $\mathscr{L}_s.$ In order to be able to assess the strength of the finite front-thickness effects, the zeroth and first-order approximations with respect to the front thickness will be considered concurrently.

\subsubsection{Critical diagrams}\label{critical}

An important and quite unexpected feature revealed by the numerical analysis is that in the region of parameters $\theta, U_f,\mathscr{L}_s$ typical of limit flames, type I solutions come in pairs having close $U$'s. Thus, for a flame with $\theta = 5,$ $U_f = 7.2$\,cm/s and $\mathscr{L}_s=-0.45$\,mm in a $9.5$\,cm diameter flammability tube, the two $U$-eigenvalues are $4.94$ and $4.65.$ There is also one type II solution which, however, has a too high speed $U = 30.7$ to be relevant to the inflammability issue; it can presumably be realized by means of a sufficiently strong ignition source and/or strong mechanical flame perturbation. For brevity, type I solutions with the lower (higher) speed will be refereed to as type Ia (type Ib). It turns out, furthermore, that despite closeness of their speeds, these solutions are essentially distinct regarding the structure of the burnt gas flow: near the inflammability limits and sufficiently small front thickness, type Ia (Ib) flames always have $\dot{Q}<0$ ($\dot{Q}>0$), that is, critical type Ia flames are metastable [Cf. Sec.~\ref{dissipation}]. Type II flames are normally characterized by large positive $\dot{Q}$'s, but dimensionless $\dot{Q}$ decreases as $U_f$ increases. A typical example is given in Table~\ref{typical}. But things may change as the front thickness increases, and in general each flame type may have $\dot{Q}$ of either sign. It can be added that regarding the transversal velocity component and the front shape, type Ia and  type Ib solutions are nearly identical.

\begin{table}
\begin{tabular}{c|cc}
\hline\hline
    solution
  & \hspace{0,5cm} $U$ \hspace{0,5cm}
  & \hspace{0,5cm} $\dot{Q}$ \hspace{0,5cm}
  \\
  type
  & (cm/s)
  &
  \\
\hline
Ia & 27.7 & $-$54.8  \\
Ib & 30.8 & 25.4  \\
II & 192  &  78580 \\
\hline\hline
\end{tabular}
\caption{Propagation speeds and rates of energy dissipation of various symmetric solutions for a flame with $\theta = 5.4,$ $U_f = 8.3$\,cm/s, $\mathscr{L}_s=0.01$ in a $9.5$\,cm wide channel.} \label{typical}
\end{table}

\begin{figure}
\includegraphics[width=0.5\textwidth]{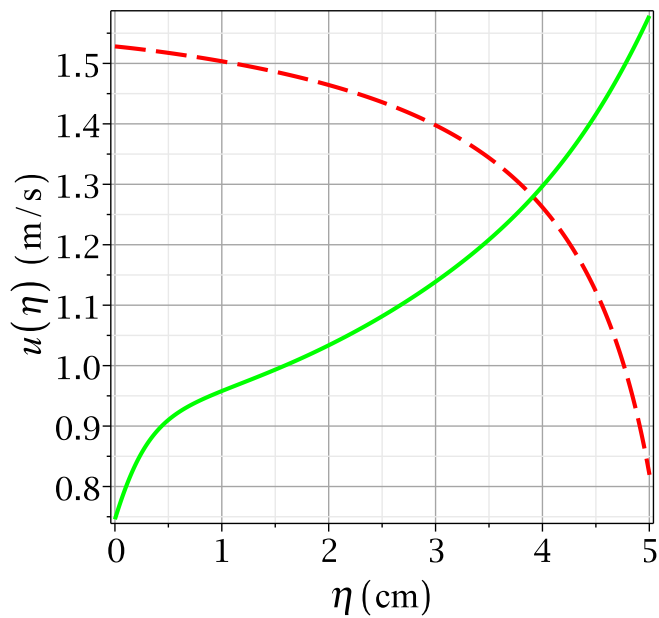}
\caption{(Color online) Burnt gas velocity distributions at the lower boundary of the transition domain for a type I flame in strong gravity (solid, $g=49$), and a flame anchored in a fast stream (dashed, $g=0$). $b=5$\,cm, $U_f=10$\,cm/s, $\theta = 4,$ $\mathscr{L}_s=0.01$.}\label{fig3}
\end{figure}

The following circumstance is the key to explaining the phenomenon of flame extinction. In the absence of gravity, the gas flow induced by the flame is normally such that the combustion products leaving the transition domain move faster near the channel centerline than near the walls. This is because, under the given overall pressure drop through the transition domain, the velocity gain is larger for smaller gas density, so that the gas elements burned near the flame tip are accelerated stronger than those traversing this domain in its upstream part. The effect of gravity [the term $2\alpha g [f(\eta) - U]$ in Eq.~(\ref{uvelocity})] is opposite: the gas elements flowing closer to the centerline are decelerated more strongly. Numerical integration of the above system shows that near the inflammability limits, the latter effect predominates over the former: under strong gravity ($g\gg 1$), the function $u(x(\eta),U)\equiv u(\eta)$ turns out to increase with $\eta,$ Fig.~\ref{fig3}. Numerical analysis reveals furthermore that type Ib solutions are singular at sufficiently small $U_f$ (that is, sufficiently large $g$). Namely, as the normal flame speed decreases to some critical $U_f$ (dependent on $\theta$), $u(\eta)$ vanishes at the channel centerline ($\eta=0$). For still smaller $U_f,$ the root $\eta_0$ of the function $u(\eta)$ shifts from the centerline towards the wall, while $u(\eta)$ becomes formally imaginary at $\eta<\eta_0,$ that is, the steady regime of flame propagation ceases to exist. Nullification of the gas velocity over a finite region behind the flame means physically that the burnt gas stops to flow out of the corresponding region of the flame front. This is, of course, impossible in a truly steady regime with a finite burning rate, but this also means that if, for some reason, a flame configuration is instantly formed with the gas velocity distribution along the front and the front shape characteristic of a supercritical type Ib solution, it will not exist longer than the transit time of burnt gas from the front to the region of vanishing velocity. In other words, the pressure distribution in this case is such that the gas burning near the channel centerline is strongly pushed upwards downstream of the front. One possible outcome of this situation is a continued essentially unsteady flame propagation, the other -- its extinction.

\begin{figure}
\includegraphics[width=0.5\textwidth]{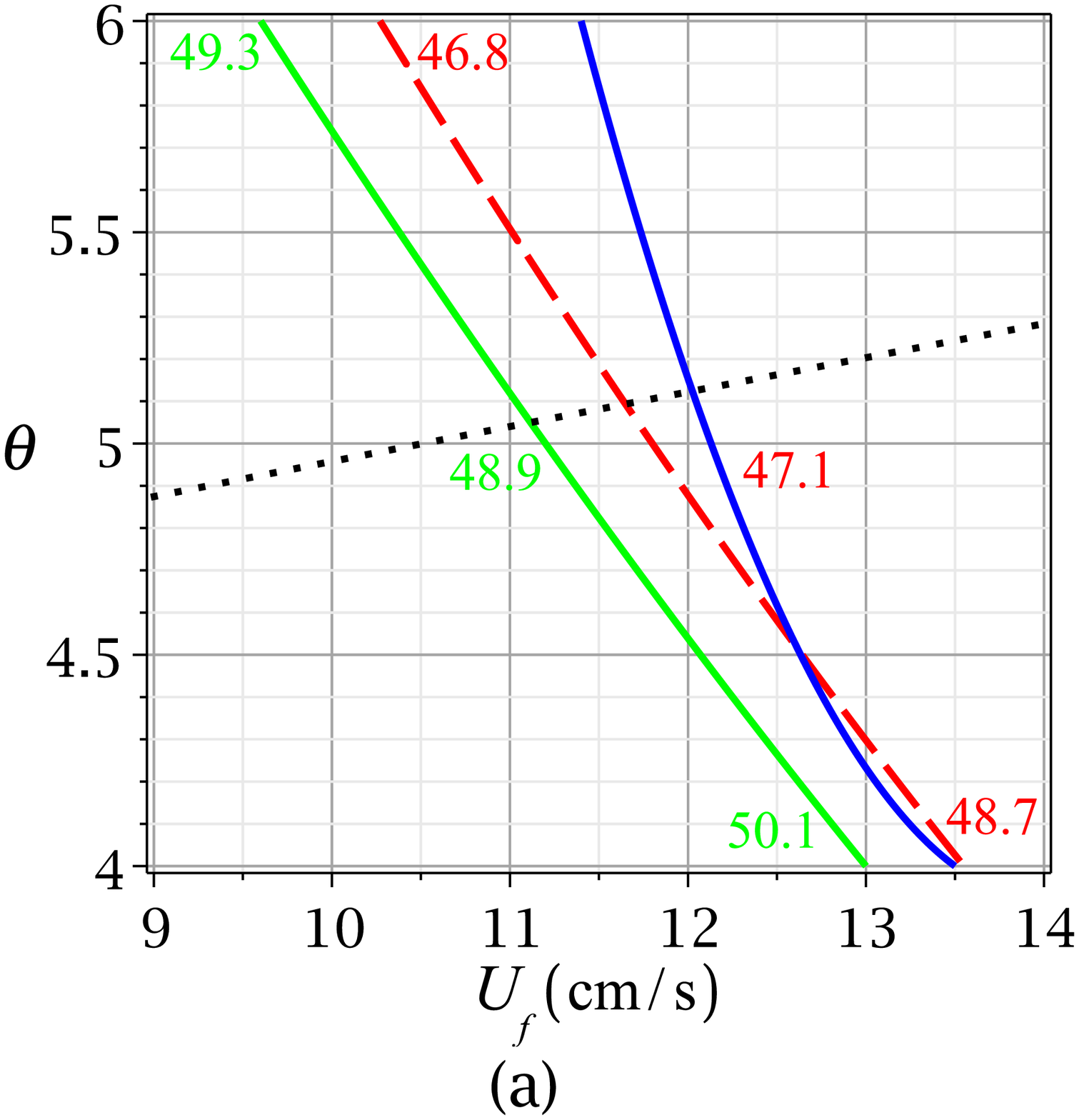}
\includegraphics[width=0.45\textwidth]{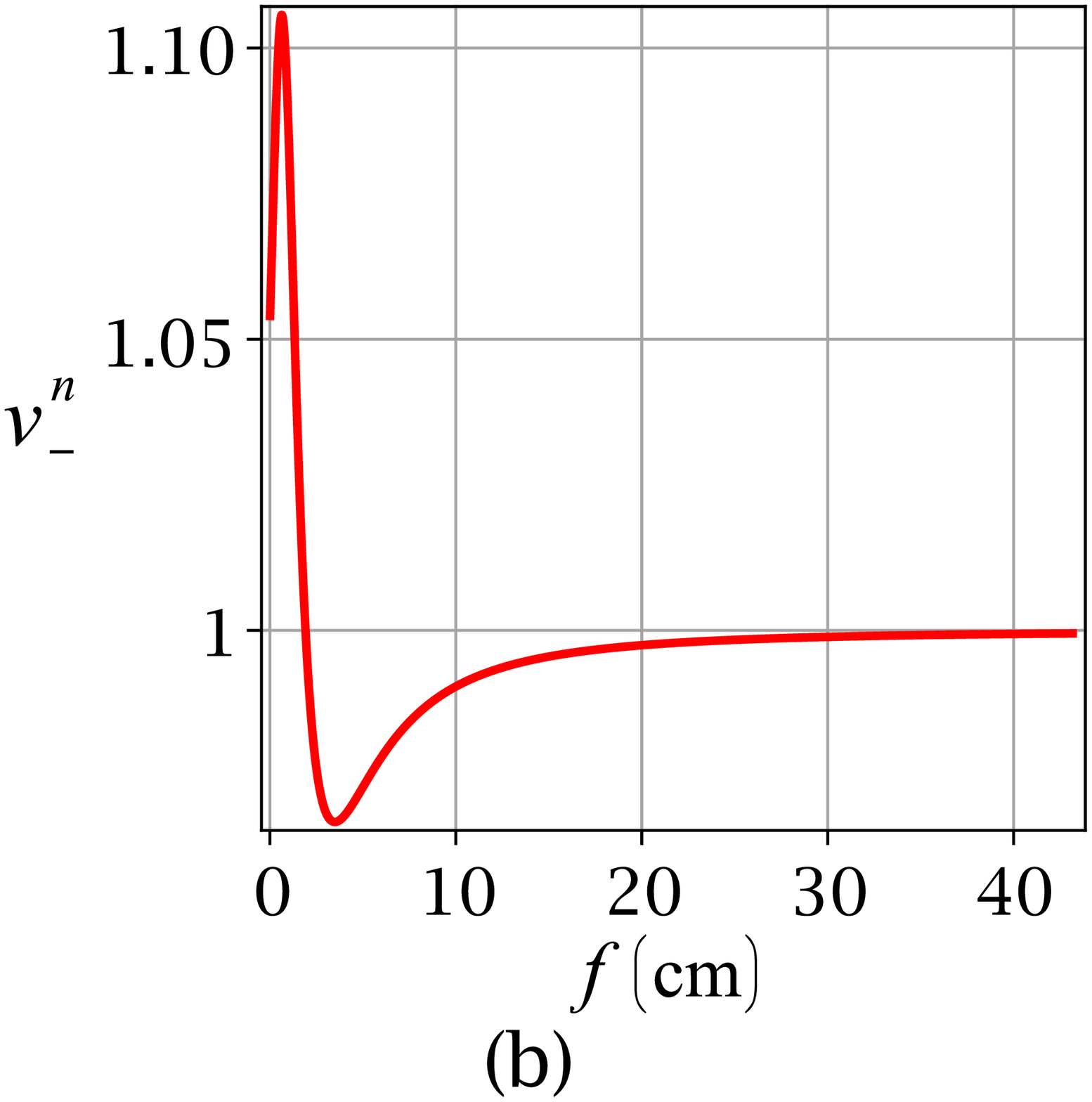}
\caption{(Color online) (a) Critical conditions for flame extinction in a $20$\,cm diameter tube. Dashed and left solid line are the curves $\eta_0=0$ for the type Ib solutions with $\mathscr{L}_s=0$ and  $\mathscr{L}_s=-0.45$\,mm, respectively; numbers illustrate variation of $U_{Ib}$ along the curves. The right solid line is the curve $\dot{Q}=0$ for type Ib solutions with $\mathscr{L}_s=-0.45$\,mm. Dotted line is the phase locus of lean methane-air flames, drawn according to Ref.~\cite{bradley} (b) Normal speed of a limit methane-air flame vs. front position. $b=20$\,cm, $U_f=11.2$\,cm/s is taken as a velocity unit.}\label{fig4}
\end{figure}

Figure~\ref{fig4}(a) is a phase diagram representing critical conditions in the $U_f$--$\theta$ plane for methane-air flames in a $20$\,cm diameter tube. The dashed and left solid lines correspond to the zeroth and first-order approximations in $l_f,$ respectively. Subcritical type Ib regimes belong to the region on the right of the critical curve, supercritical -- on the left. The right solid line is the curve $\dot{Q}=0$ for type Ib solutions; they are metastable to the right of this curve. $U_f$ for near-limit flames is not easy to determine accurately, but the propagation speed of type Ib flames turns out to change only a little along the critical curves, when expressed in centimeters per second. The three numbers near each curve give this speed at the ordinates $\theta = 4, 5$ and $6.$ The value\cite{bradley} $\mathscr{L}_s = -0.45$\,mm is taken for the methane-air flame near the lean inflammability limit in the $20$\,cm diameter tube (equivalence ratio $\phi\approx 0.6$). In units of the channel half-width, this is $\mathscr{L}_s = -0.0045.$ The fact that the relative increase in $U$ brought about by the first-order correction is some ten times as large reveals extreme susceptibility of the near-limit flames to the finite front-thickness effects. This property turns out to be of global dynamical origin. Namely, numerical analysis shows that if the finite front-thickness corrections are included only in the local propagation law (the evolution equation), but not in the dynamical equations (the flow equations), the shift of the critical curve as well as the flame speed variation along it are nearly negligible, despite significant changes in the normal flame speed (fractional correction to the normal flame speed peaks near the flame tip, exceeding there $10\%,$ Fig~\ref{fig4}(b)). This is readily understood once it is observed that the terms proportional to $\mathscr{L}_s$ are directly coupled to gravity only in the last term of the master equation (\ref{master3}). On the other hand, their indirect coupling through the terms involving the function $x(f),$ the form (\ref{x(f)}) of which is determined by the evolution equation, is strongly suppressed because the normal speed is significantly affected only near the flame tip, where $x(f)\approx 0.$

The calculated flame propagation speed is to be compared with the measured in Ref.~\cite{vonlavante}, according to which the observed $U$ fluctuated between $50$ and $60$\,cm/s. As the authors indicate, the flame very rarely had a regular shape, and cellular front structure was developed. Cells augment the front area, hence, the flame propagation speed, therefore, the lowest measured value is to be taken for comparison, that is $50$\,cm/s. For each pair fuel/oxidyser, both $U_f$ and $\theta$ change with the mixture composition, and the dotted line in Fig.~\ref{fig4}(a) is the corresponding trajectory for lean methane-air flames, drawn using the results of Ref.~\cite{bradley} It is seen that the critical flame in a $20$\,cm diameter tube has $U=48.9$\,cm/s. As discussed in the Introduction, the error of the asymptotic calculation is difficult to determine because no estimate of the remainder of the expansion in $l_f$ exists. Part of the calculational error related to the large-slope expansion is about $10\%,$ according to Ref.~\cite{kazakov3}

\begin{figure}
\includegraphics[width=0.49\textwidth]{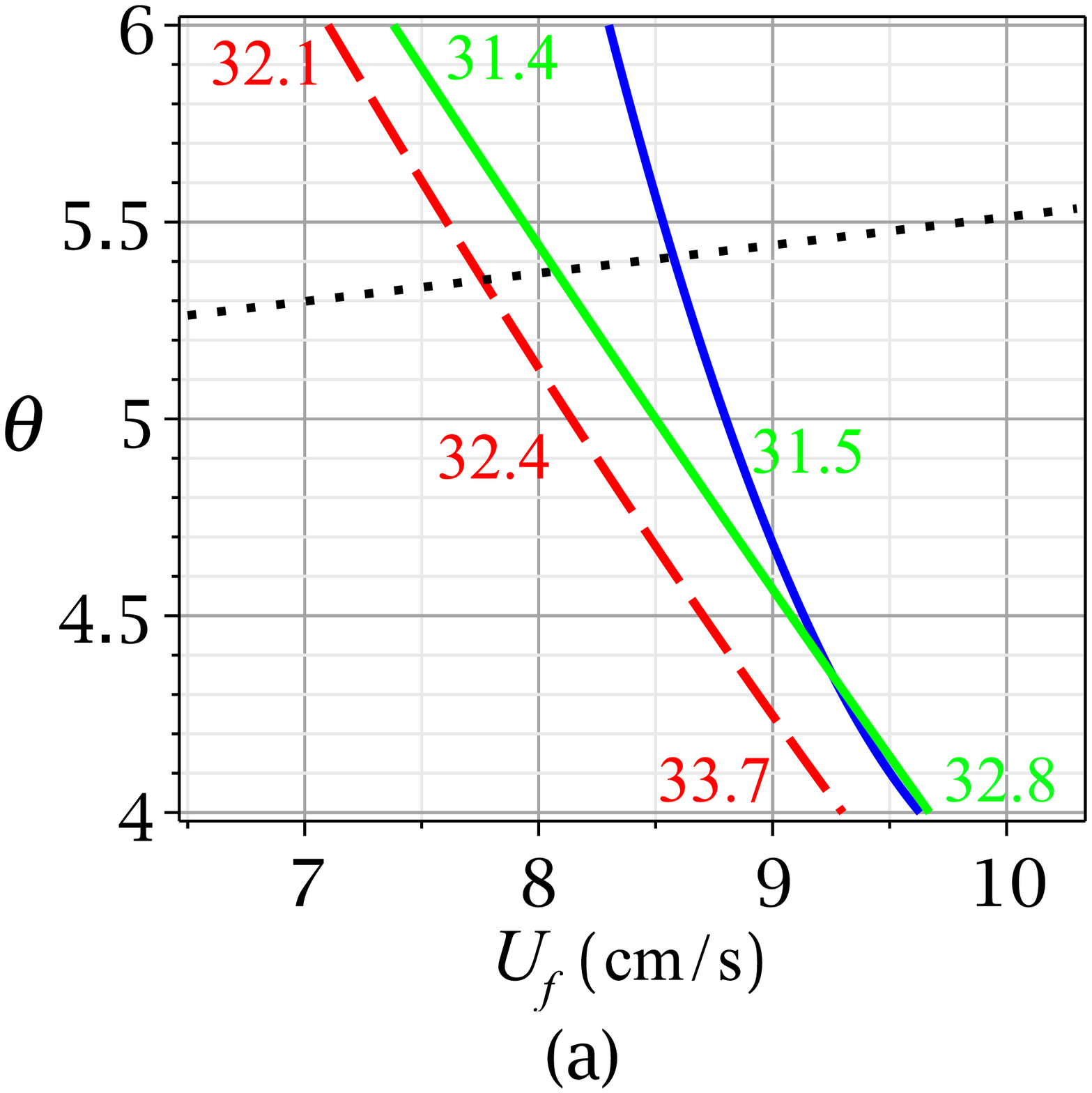}
\includegraphics[width=0.49\textwidth]{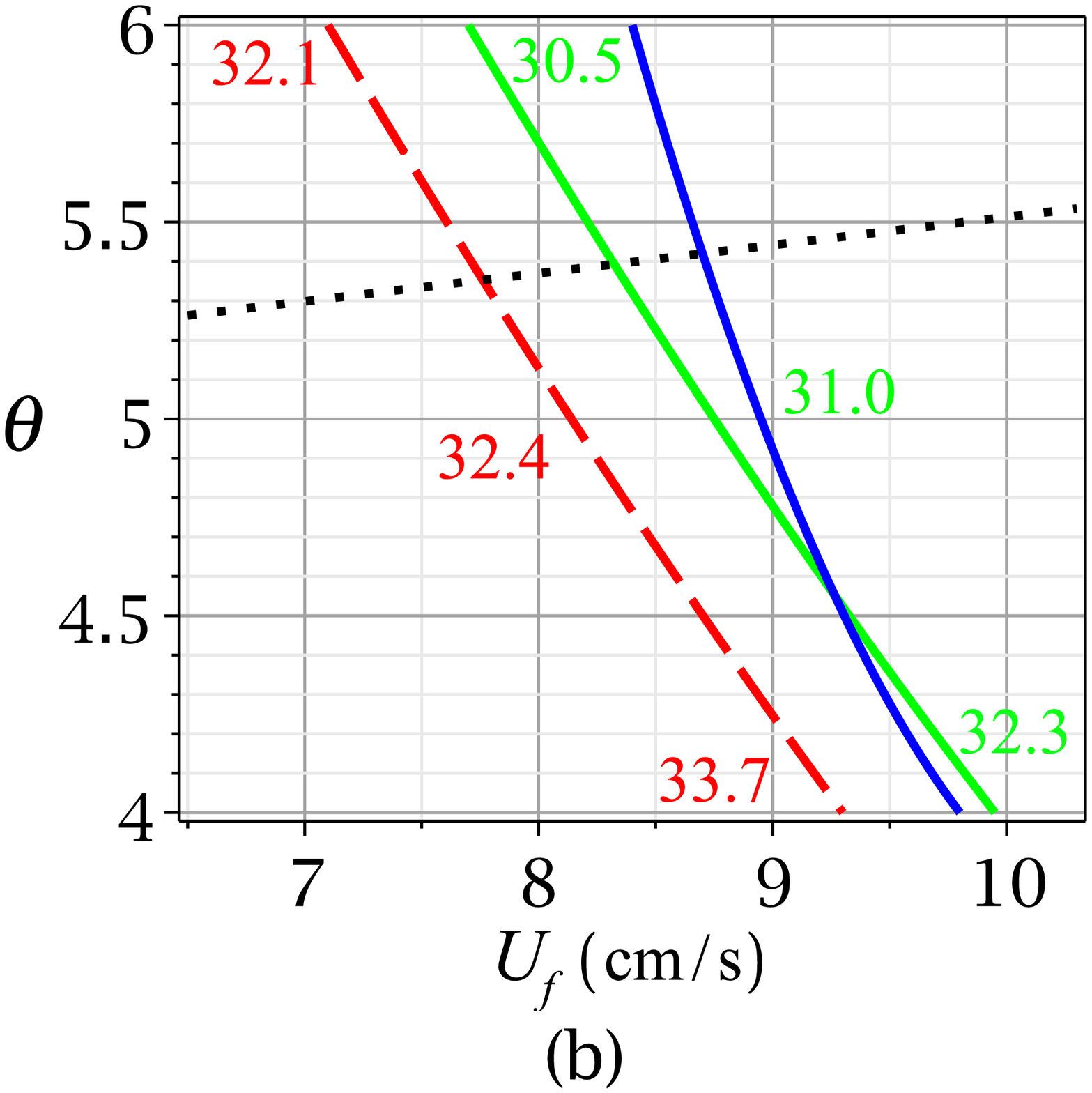}
\caption{(Color online) Critical diagrams for flames with $\mathscr{L}_s=0.25$\,mm (a) and  $\mathscr{L}_s=0.5$\,mm (b) in a $9.5$\,cm-diameter tube. Same notation as in Fig.~\ref{fig4}. The phase locus of lean propane-air flames (dotted line) is drawn according to Refs.~\cite{taylor,calc}}\label{fig5}
\end{figure}

\begin{figure}
\includegraphics[width=0.5\textwidth]{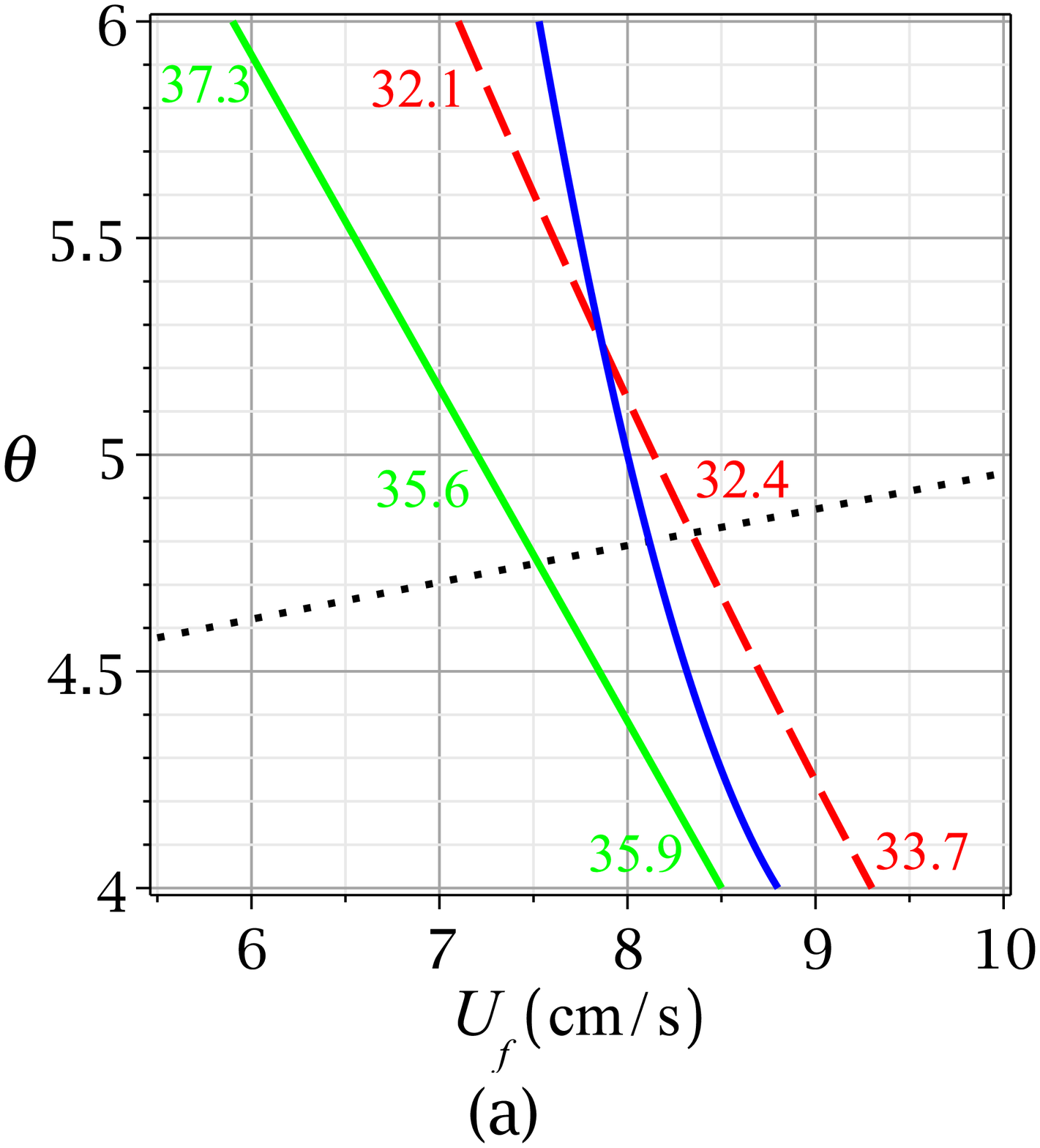}
\includegraphics[width=0.49\textwidth]{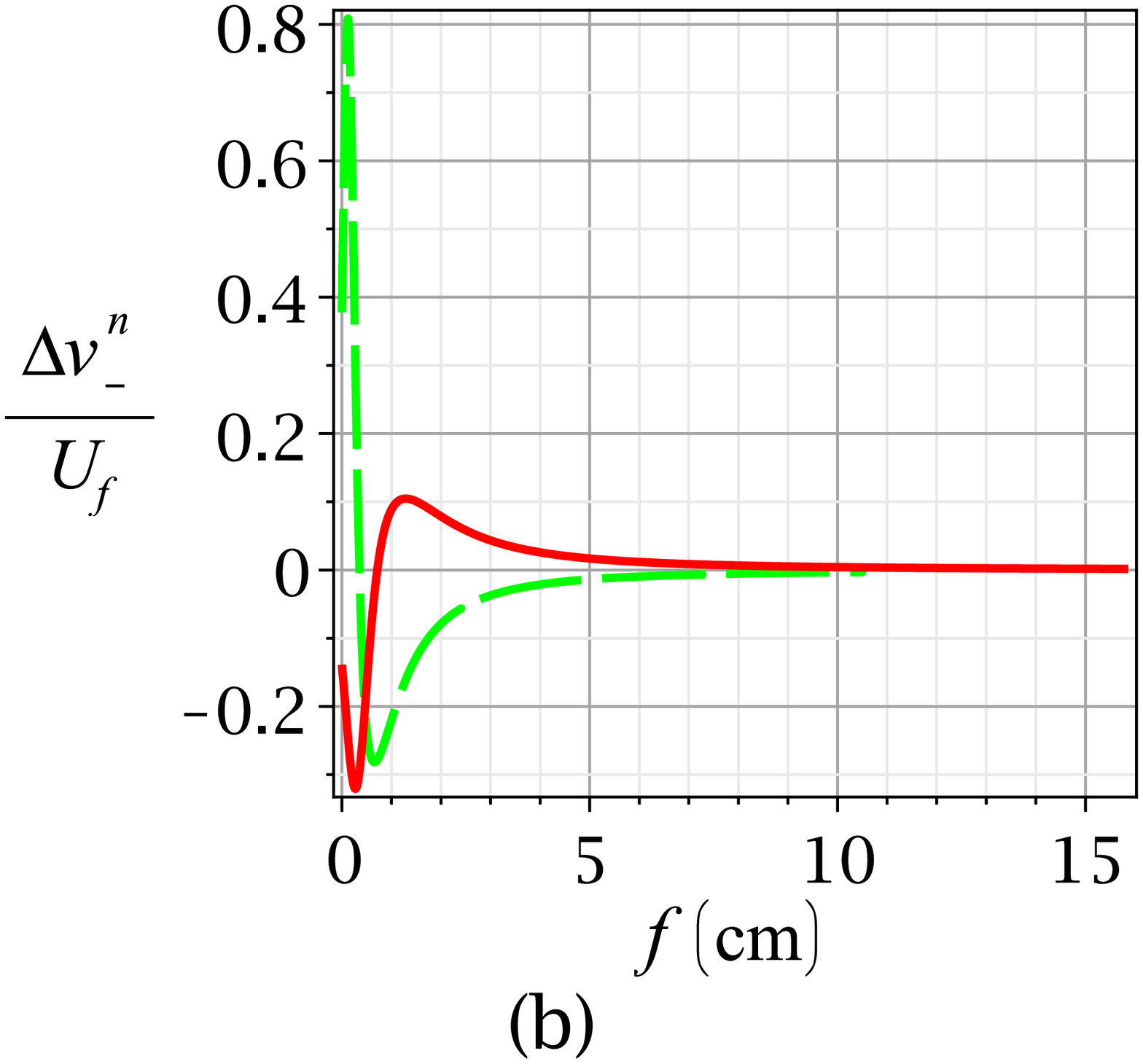}
\caption{(Color online) (a) Critical diagram for methane-air flames in a $9.5$\,cm-diameter tube. Same notation as in Fig.~\ref{fig4}. (b) Fractional correction to the normal speed of a limit flame in a $9.5$\,cm-diameter tube (solid line, $\theta=5.4,$ $U_f=8.3$\,cm/s, $\mathscr{L}_s=0.5$\,mm), and calculated on a zero-order type I solution in a $5.1$\,cm-diameter tube (dashed line, $\theta = 6,$ $U_f = 5.2$\,cm/s, $\mathscr{L}_s=-0.62$\,mm).}\label{fig6}
\end{figure}

Turning to the cases $b=9.5$\,cm and $b=10$\,cm, the available experimental data\cite{levy1965,vonlavante} reproduced in Table~\ref{table1} is spread significantly less than for $b=20$\,cm. It allows comparison for flames with $\mathscr{L}_s$ positive (propane-air) as well as negative (methane-air). As $\mathscr{L}_s$ is not known for propane-air mixtures, Figs.~\ref{fig5}(a) and \ref{fig5}(b) give critical conditions for two different positive $\mathscr{L}_s$ ($0.25$\,mm and $0.5$\,mm), along with those for a zero-thickness flame. Figure~\ref{fig6}(a) is for methane-air flames in a tube with $b=9.5$ (the difference between diagrams for $b=9.5$\,cm and $b=10$\,cm is negligible compared to the experimental error). It is seen that as $\mathscr{L}_s$ increases, the critical curves shift rightwards, while the flame propagation speed decreases. Incidentally, this conclusion concurs with the experimental evidence\cite{levy1965} suggesting that the limit propane-air flames are somewhat slower than methane-air flames in tubes of the same diameter. It is to be noted that exactly the opposite would take place if the limit flames were driven by buoyancy, as sometimes supposed in view of the observed similarity of their shape and speed with those of bubbles. Indeed, both the local burning rate of propane-air flames and their gas expansion coefficient are larger than those of methane-air flames. But the larger $U_f$ and $\theta,$ the larger the total burning rate and the buoyancy force. Therefore, the flame propagation speed ought to increase with $U_f,\theta.$ It is worth reemphasizing in this connection that flame propagation in a strong gravitational field is driven not by buoyancy, which is a term proper to separating surfaces of a quite different type, but by the baroclinic effect -- a gravity-induced vorticity production in the flame front. It is the balance of baroclinic vorticity with that generated by the front curvature which determines the flame structure and its propagation speed.

Also noteworthy is the appearance of an intersection of the critical curve for propane-air flames with the curve $\dot{Q}=0.$ As $\mathscr{L}_s$ increases, the crosspoint shifts up along the curve. For sufficiently large $\mathscr{L}_s,$ it might go above the point representing critical conditions in the given mixture (which corresponds to the intersection of the critical curve with the dotted line). This would mean loss of a truly stable regime of flame propagation: the flame would be periodically disturbed by the burnt gas flow rearrangements conditioned by the flow metastability. Since $\mathscr{L}_s$ is not known, it is impossible to say whether this possibility is realized for propane-air flames, but if it is, its description is again beyond the capabilities of the small-$l_f$ expansion. In fact, Fig.~\ref{fig6}(b) shows that the finite front-thickness correction to the normal flame speed is not really small already for $\mathscr{L}_s = 0.01.$ The results of comparison of the theory with experiment are summarized in Table~\ref{table1} (wherein $\mathscr{L}_s$ for propane-air flame is taken equal $0.25$\,mm, as in Fig.~\ref{fig5}(a)).

\begin{table}
\begin{tabular}{c|ccc|ccc}
\hline\hline
  \hspace{0,1cm} \hspace{0,2cm}
  & \hspace{0,2cm} $\theta$ \hspace{0,2cm}
  & \hspace{0,3cm} $U_f$ \hspace{0,3cm}
  & \hspace{0,3cm} $2\mathscr{L}_s/b$ \hspace{0,3cm}
  & \hspace{0,3cm} $U_{Ia}$ \hspace{0,3cm}
  & \hspace{0,3cm} $U_{Ib}$ \hspace{0,3cm}
  & $U_{exp}$
  \\
  flame
  &
  & \hspace{0,3cm} (cm/s) \hspace{0,3cm}
  &
  &\hspace{0,3cm} (cm/s) \hspace{0,3cm}
  &\hspace{0,3cm} (cm/s) \hspace{0,3cm}
  &\hspace{0,3cm} (cm/s) \hspace{0,3cm}
    \\
\hline
propane-air & 5.4 & 8.3 & 0.005  & 28.2 & 31.1 & 31.7 $\pm$ 1.6 \\
methane-air & 4.75 & 7.5 & $-$0.01 & 33.7 & 35.6 & $\left\{
\begin{array}{cc}
33.1\pm 2.3^{[3]} \\
33.5\pm 1.5^{[8]}
\end{array}
\right.
$ \\
\hline\hline
\end{tabular}
\caption{Methane-air and propane-air flames in the lean inflammability limit. $\theta$ and $U_f$ correspond to the intersections of the critical curves in Figs.~\ref{fig5}(a), \ref{fig6}(a) with the dotted lines. $U_{exp}$ is the measured flame propagation speed, $U_{Ia}$ and $U_{Ib}$ its values as given by the type Ia and type Ib solutions for $b=9.5$\,cm. The superscripts [3], [8] are respectively for Ref.~\cite{levy1965} (circular tube with $b=9.5$\,cm) and Ref.~\cite{vonlavante} (square tube with $b = 10$\,cm), $\pm$ symbolize the standard deviation.} \label{table1}
\end{table}

\begin{figure}
\includegraphics[width=0.5\textwidth]{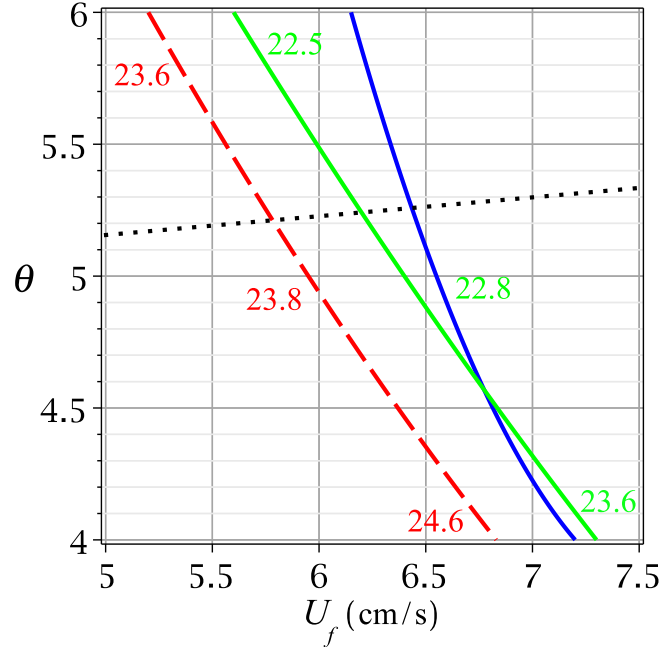}
\caption{(Color online) Critical diagram for propane-air flames in a $5.1$\,cm-diameter tube. Same notation as in Fig.~\ref{fig4}. }\label{fig7}
\end{figure}

Finally, let us consider flame propagation in tubes with $b=5.1$\,cm. The critical diagram for propane-air flames with $\mathscr{L}_s = 0.25$\,mm, Fig.~\ref{fig7}, is of the same structure as before. In contrast, it is entirely changed by the finite front-thickness correction in the case of methane-air flames. Specifically, all type Ia solutions become stable on account of this correction, so that the curve $\eta_0=0$ no longer has the meaning of a critical condition for flame extinction (since the lean inflammability limit lowers as $b$ decreases, the Markstein length for flames under consideration is somewhat larger in magnitude,\cite{bradley} $\mathscr{L}_s = -0.62$\,mm ($\phi\approx 0.55$)). However, this dramatic change in the properties of solutions occurs in a situation where the finite front-thickness corrections in no way can be considered small. In fact, the fractional correction to the normal flame speed becomes comparable to unity, Fig.~\ref{fig6}(b), while the curve $u(0)=0$ moves to the region of very small normal speeds, $U_f\approx 2-3$\,cm/s (it would be out of the field of view in Fig.~\ref{fig7}). This evidently means that the first-order approximation in $l_f$ is no longer applicable in this case. At the same time, Table~\ref{table2} indicates that the zero-thickness approximation still works well. This curious fact deserves further discussion. There is little doubt that the breakdown of the small-$l_f$ expansion for methane-air flames in a $5.1$\,cm diameter tube is the result of a changeover between the gravitational length scale $U^2_f/g$ and $\mathscr{L}_s.$ Indeed, comparison of Figs.~\ref{fig4}--\ref{fig7} shows that the critical normal speeds drop as the tube diameter decreases. In addition to that, critical curves for methane-air flames shift {\it leftwards} from those for zero-thickness flames. The scale $U^2_f/g$ for critical methane-air flames thus drops even faster, and in the case under consideration turns out to be less than $\mathscr{L}_s$ ($\mathscr{L}_s g/U^2_f \gtrsim 1.5$), making the small-$l_f$ expansion inapplicable.

\begin{table}
\begin{tabular}{c|ccc|ccc}
\hline\hline
  \hspace{0,1cm} \hspace{0,2cm}
  & \hspace{0,2cm} $\theta$ \hspace{0,2cm}
  & \hspace{0,3cm} $U_f$ \hspace{0,3cm}
  & \hspace{0,3cm} $2\mathscr{L}_s/b$ \hspace{0,3cm}
  & \hspace{0,3cm} $U_{Ia}$ \hspace{0,3cm}
  & \hspace{0,3cm} $U_{Ib}$ \hspace{0,3cm}
  & $U_{exp}$
  \\
  flame
  &
  & \hspace{0,3cm} (cm/s) \hspace{0,3cm}
  &
  &\hspace{0,3cm} (cm/s) \hspace{0,3cm}
  &\hspace{0,3cm} (cm/s) \hspace{0,3cm}
  &\hspace{0,3cm} (cm/s) \hspace{0,3cm}
    \\
\hline
propane-air & 5.25 & 6.2 & 0.01  & 20.4 & 22.6 & 23.1 $\pm$ 1.3 \\
methane-air & 4.65 & 6.3 & 0 & 22.3 & 24.0 & $\left\{
\begin{array}{cc}
23.5\pm 0.6^{[3]} \\
22.8\pm 0.5^{[8]}
\end{array}
\right.
$ \\
\hline\hline
\end{tabular}
\caption{Same for $b=5.1$\,cm.} \label{table2}
\end{table}

\subsubsection{Dynamics of partial flame propagation and extinction}\label{dynamics}

The whole picture of partial flame propagation can now be described as follows. Ignition by means of a weak energy source results in a steady flame with the lowest propagation speed, that is type Ia flame. The flame travels as such for some distance, but since it is characterized by $\dot{Q}<0,$ sooner or later it will have to go over to a regime with $\dot{Q}>0.$ The nearest solution with $\dot{Q}>0$ is the slightly faster type Ib regime. If the latter is subcritical, the flame continues to propagate with a somewhat increased speed, which is thus the flame speed that will be measured in a sufficiently long tube. On the other hand, if the flame is supercritical, then as was already mentioned, flame extinction following a short transient is one of the two possible results of the loss of flame stationarity. It is to be expected for flames less energetic (leaner), hence less resistant to the flow perturbation taking place near the flame centre. Since the critical curves in Fig.~\ref{fig3} shift to the leaner side as the tube diameter decreases, this outcome is favored in sufficiently narrow tubes, and as the experiments show, it is actually observed in tubes with $b\lesssim 10$\,cm. In wider tubes, on the other hand, the flame might continue to propagate, but in an essentially unsteady regime, {\it e.g.}, break into cells.

Characteristics (i)--(iv) of the observed flame behavior are readily understood in this picture. Figures~\ref{fig4}--\ref{fig7} demonstrate that the critical normal flame speed increases with the tube diameter, that is, the inflammability range narrows in wider tubes, in agreement with the observation (ii). The existence of the type Ia regime under  critical conditions makes possible comparatively long steady flame propagation before extinction [the first part of (i)], whereas transition to the slightly faster short-lived type Ib flame configuration accounts for the rest of (i). It was already discussed how specifics of the type Ib solution explain the fact that extinction begins at the flame centre [the last observation in (iv)]. In this connection, a question may arise as to whether flame extinction could be partial, that is, after its central part has extinguished, the flame could propagate filling only part of the tube. The following simple consideration shows that this is impossible. One first notes that the vanishing of the flame centre produces the same effect as a reduction of the tube cross-section, which is particularly evident in the two-dimensional setting. Since the speed $U_{Ib}$ of a critical flame lowers as the channel width decreases, the propagation speed of the flame under consideration turns out to be larger than the speed of a critical flame. At the same time, $\eta_0$ grows with $U,$ and as calculations show, does so very rapidly. For example, in the case of a critical propane-air flame in a $9.5$\,cm-diameter tube ($\theta = 5.4,$ $U_f=8.3$\,cm/s, $\mathscr{L}_s = 0.01$), $\eta_0=0$ for $U_{Ib}=3.71,$ but it is as large as $2.4$\,cm already for $U=3.9.$ An increased $\eta_0$ effectively reduces the channel width further, which in turn increases the difference between given $U$ and $U_{Ib}$ of a critical flame. The process of flame extinction is thus self-accelerating, and its characteristic time is quite short -- it is of the order of the type Ib flame lifetime; to quote the author of Ref.~\cite{levy1965}, the flame disappears from view as if rising past an opaque screen. Turning finally to (iii), it should now be clear that the same property of type Ib solutions establishes a direct link between limit flames and bubbles: the flow stoppage developing from the flame centre towards the walls makes the two structures alike. A quantitative comparison of the theory with experiment has already been given in Tables~\ref{table1}, \ref{table2}.

Regarding the similarity of limit flames with bubbles, its transitory nature should be emphasized. Although the burnt gas slowdown takes place in both Ia and Ib regimes, the gas velocity in type Ia solutions remains large compared to $\theta U_f$ everywhere in the channel cross-section at $y=U,$ so that no analogy with bubbles exists. This analogy emerges only at the latest stage of flame evolution following transition to a critical type Ib regime, when the burnt gas stops to flow along the centerline, resulting in the local flame extinction which then rapidly spreads out along the front. In other words, steady flame never behaves like a bubble, but so does the hot postflame structure after extinction, that is the type Ib flame remnant.

The fact that the bubble analogy holds only under the critical conditions has another interesting implication. The bubble speed is known to scale with $g,b$ as $U\sim \sqrt{gb}.$ At the same time, it is not difficult to see that $g$ cannot be eliminated from Eqs.~(\ref{master3}), (\ref{3rel}) by rescaling $u_- \to \sqrt{g}\,u_-,$ $f'\to \sqrt{g}f'.$ This means that the dependence of $U$ on $g$ is more complicated in general. Thus, the scaling $U \sim \sqrt{gb}$ holds only on the critical solutions, and as the numerical analysis shows, it does so only approximately.

\subsection{Flame acceleration in smooth tubes}\label{acceleraion}

\subsubsection{Acceleration of methane-air flames in an open vertical tube}\label{openaccel}

Opening the upper end of the tube makes it a ``chimney'': the hydrostatic pressure difference of the ambient cold air forces the inner gases to accelerate. This adds to the acceleration generated by the difference of dynamic gas pressure at the tube ends, which exists already in horizontal tubes.\cite{kazakov4} It is not difficult to relate these quantities. Denoting flame acceleration in the laboratory frame by $a$ (defined positive if the flame accelerates upwards), an equivalent gravity in the reference frame attached to the front is equal in value and opposite to $a,$ so that the gravitational potential is now $\phi = - (g+a)y.$ Combining Eqs.~(\ref{momentumup}), (\ref{momentumdown}), (\ref{momentuminf}) and using the jump conditions at the front as before readily gives
\begin{eqnarray}
\tilde{p}|_{y=H} - \tilde{p}|_{y=0} + \alpha U^2 = \frac{\alpha (g+a)}{\theta}\int_{0}^{1}d\eta f(\eta).
\end{eqnarray}
\noindent By virtue of the flow homogeneity upstream of the transition domain, $\tilde{p}|_{y=0}$ is equal to its value at the upper tube end. Assuming also that the distance $s$ travelled by the flame from the lower end is large enough to ensure homogeneity of the burnt gas flow at the  exit, so that one can set $H=s,$ and substituting $(p|_{y=s} - p|_{y=s-L}) = gL,$ where $L$ is the tube length, one finds
\begin{eqnarray}\label{accel}
g+a = (\alpha U^2 + gL)\left\{L - \frac{\alpha}{\theta}s + \frac{\alpha}{\theta}\int_{0}^{1}d\eta f(\eta)\right\}^{-1}.
\end{eqnarray}
\noindent Pressure drop due to the hydrodynamic resistance of the tube would reduce the numerator in this formula, but it is negligible under conditions considered below.\cite{kazakov4} Validity of the consideration using the steady equations requires sufficient slowness of the changes in $U$ and $a.$ These quasi-steady conditions read
\begin{eqnarray}\label{quasisteady}
\frac{dU}{ds}\,V\ll U, \quad \frac{da}{ds}\,V\ll a,
\end{eqnarray}
\noindent where $V=ds/dt$ is the flame speed in the laboratory frame. They ensure smallness of the fractional changes in $U,a$ during the travel of gases through the transition domain, which is sufficient to guarantee quasi-steadiness of the whole process.\cite{kazakov4} When applied to $a$ given by Eq.~(\ref{accel}), these conditions imply the following bound on its value
\begin{eqnarray}
a \ll \frac{\alpha U^2 + gL}{V},\nonumber
\end{eqnarray}
\noindent or in the ordinary units,
\begin{eqnarray}\label{quasisteady1}
a \ll (\alpha U^2 + gL)\frac{U_f}{Vb}.
\end{eqnarray}
\noindent Since $a\to 0$ as $L\to \infty,$ this inequality holds in sufficiently long tubes (notice that its strength is nonuniform -- it weakens on approaching the upper tube end). It is particularly useful and in most cases is sufficient to ensure fulfilment of the quasi-steady conditions (a notable exception will be encountered later on in this section).

\begin{figure}
\includegraphics[width=0.7\textwidth]{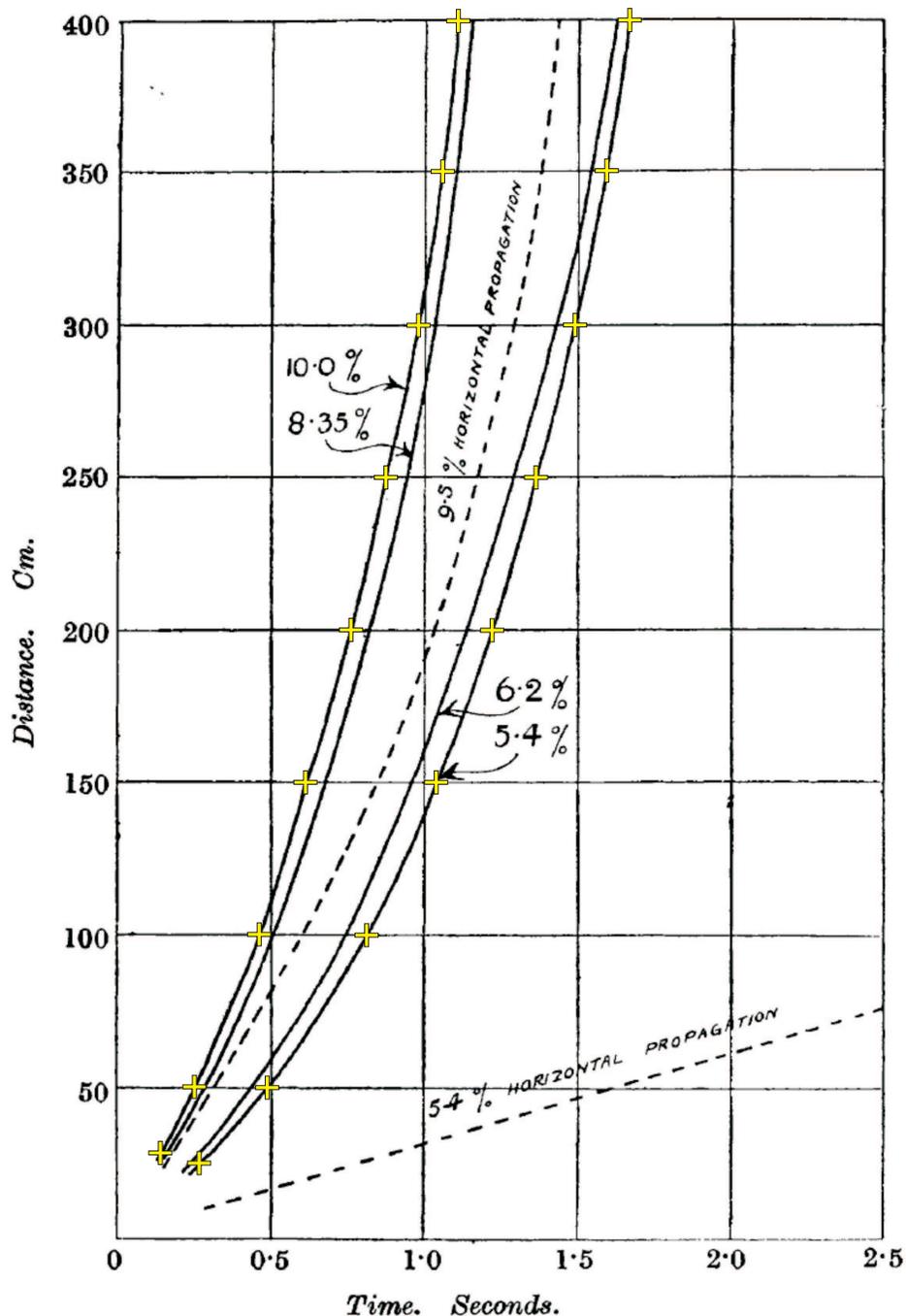}
\caption{Trajectories of various methane-air flames propagating in a vertical tube with $d=5$\,cm, $L=5$\,m. Marks on the $5.4\%$ and $9.5\%$ curves are the basis points used to construct approximating polynomials. {\it Source}: Fig.~3 of Ref.~\cite{mason1920} -- Reproduced by permission of The Royal Society of Chemistry.}\label{fig8}
\end{figure}

\begin{figure}
\includegraphics[width=0.45\textwidth]{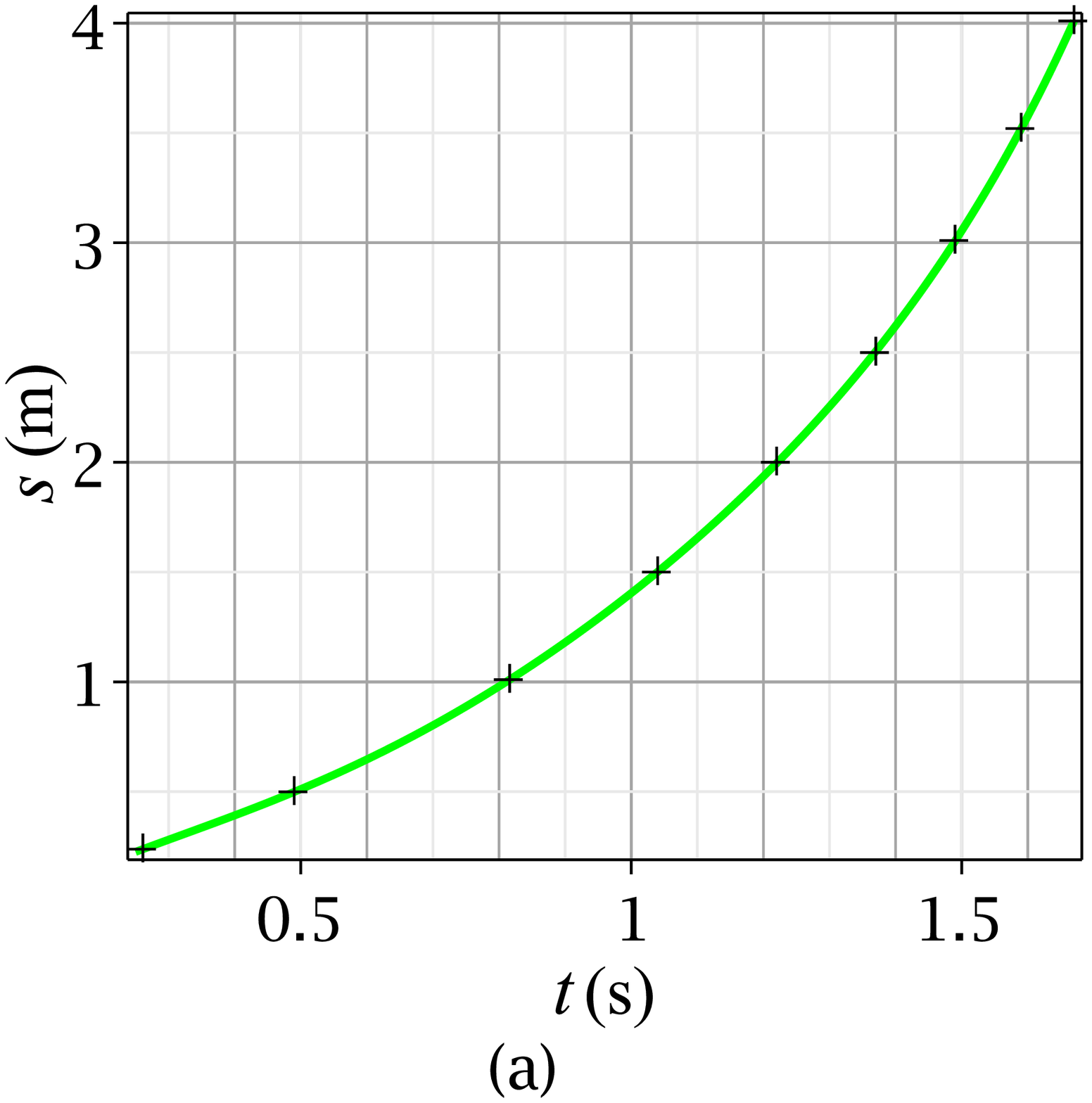}
\includegraphics[width=0.45\textwidth]{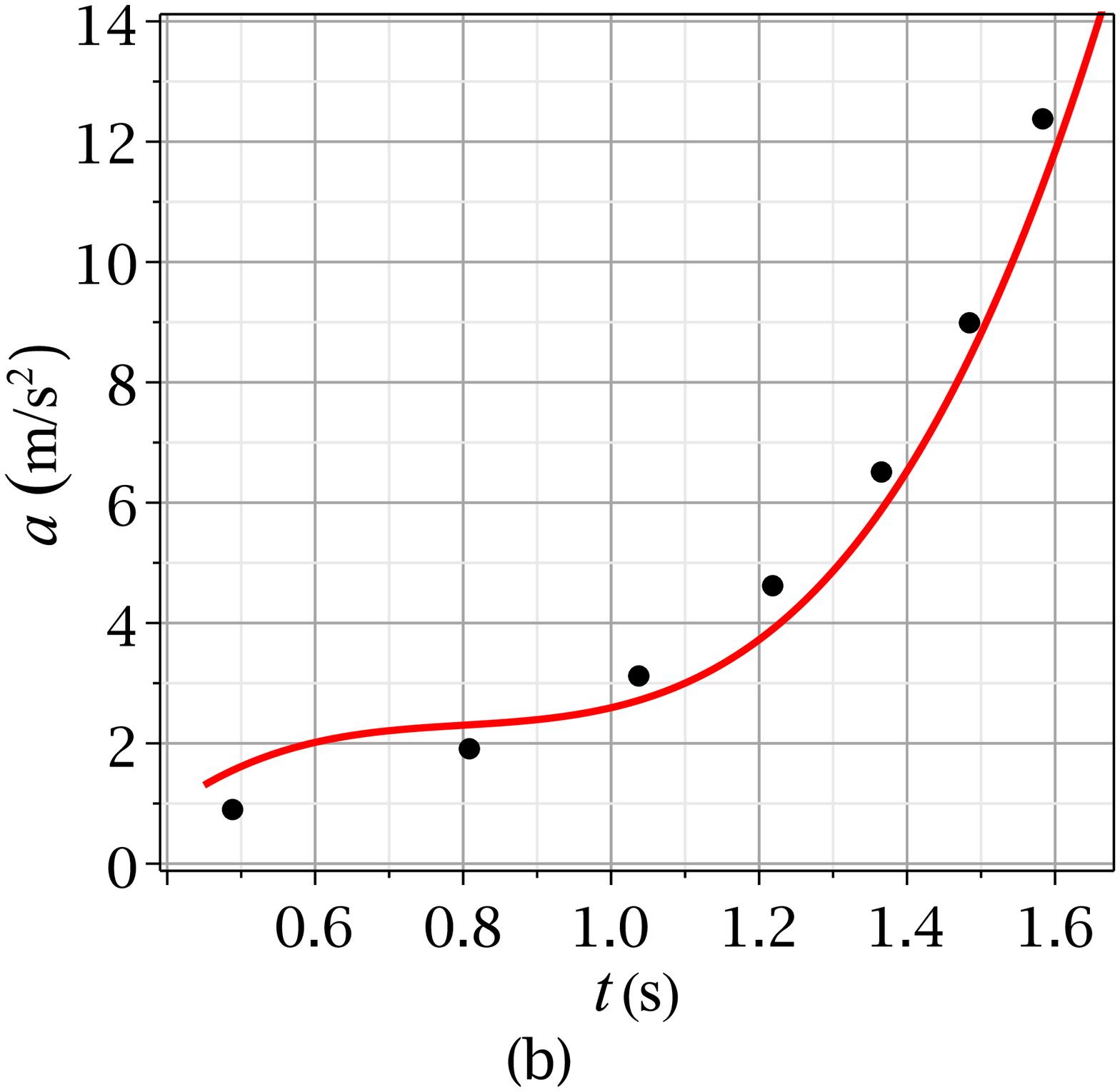}
\caption{(Color online) (a) Approximation of the $5.4\%$-methane flame trajectory (solid line), obtained by interpolating the basis points (marks) translated from Fig.~\ref{fig8}. The interpolating polynomial $S(t) = -.1397+2.051\, t-3.57\, t^2+5.662\, t^3-3.467\, t^4+.8682\, t^5$ gives the flame position in meters at the time instant $t$ expressed in seconds. (b) Flame acceleration found as the second derivative of the polynomial $S(t)$ (solid line), and as eigenvalues of type I solutions at selected $s$ (marks).}\label{fig9}
\end{figure}

The results of a classic experiment\cite{mason1920} on methane-air flame acceleration in a $5$\,cm diameter $5$\,m long tube are reproduced in Fig.~\ref{fig8}. Shown is the flame position within the tube versus time for various methane concentrations. Processing of the data was described in detail elsewhere;\cite{kazakov4} it consists essentially in polynomial approximation of the position curves using a set of basis points (crosses) digitized with the help of the DigitizeIt software.\cite{digitizeit} The second derivative of a polynomial can then be compared to the values of flame acceleration found numerically as part of the type I or type II solutions. Variety of the propagation regimes is now significantly wider than in the horizontal case. This is because of the existence of different physical solutions of each type, and the two possible realizations of each regime -- symmetric and asymmetric. Things are further complicated by the necessity to take into account heat losses in the asymmetric configuration, which are difficult to determine accurately (heat losses in a $5$\,cm diameter tube have negligible effect on the propagation of axisymmetric flames\cite{vonlavante}). Fortunately, there are two important cases which are free of these complications. First, as discussed in Sec.~\ref{inflammability}, the near-limit flame in a $5.4\%$-methane mixture is symmetric, so that the heat losses are immaterial. On the theoretical side, although there are two type I regimes of propagation, it is seen from Eq.~(\ref{accel}) that their accelerations are nearly the same, because the term $\alpha U^2$ in this case is very small compared to the gravitational contribution. On the other hand, the $U$-eigenvalue of the type II solution is so large compared to those of type I (at $s=0.5$\,m, for instance, $U_{II} = 170$\,cm/s, whereas $U_{Ib} = 24$\,cm/s) that the question which of them was observed is never raised. The calculated and measured accelerations are plotted against time in Fig.~\ref{fig9}, where the polynomial interpolation of the basis points is also shown. The error in $a(t)$ obtained by differentiating the polynomial is about $20\%$ (this figure bounds the relative change in $a$ brought in by shifts of the basis points within the thickness of the experimental curve in Fig.~\ref{fig8}, subject to the requirement that they leave the function $a(t)$ monotonic\cite{kazakov4}). As to the quasi-steady condition, it is satisfied reasonably well over most of the observed flame travel: the ratio, $r,$ of the left- and right-hand sides of (\ref{quasisteady1}) is only $0.035$ at $s=1$\,m, reaching $r=0.32$ at $s=3$\,m.

\begin{figure}
\includegraphics[width=0.45\textwidth]{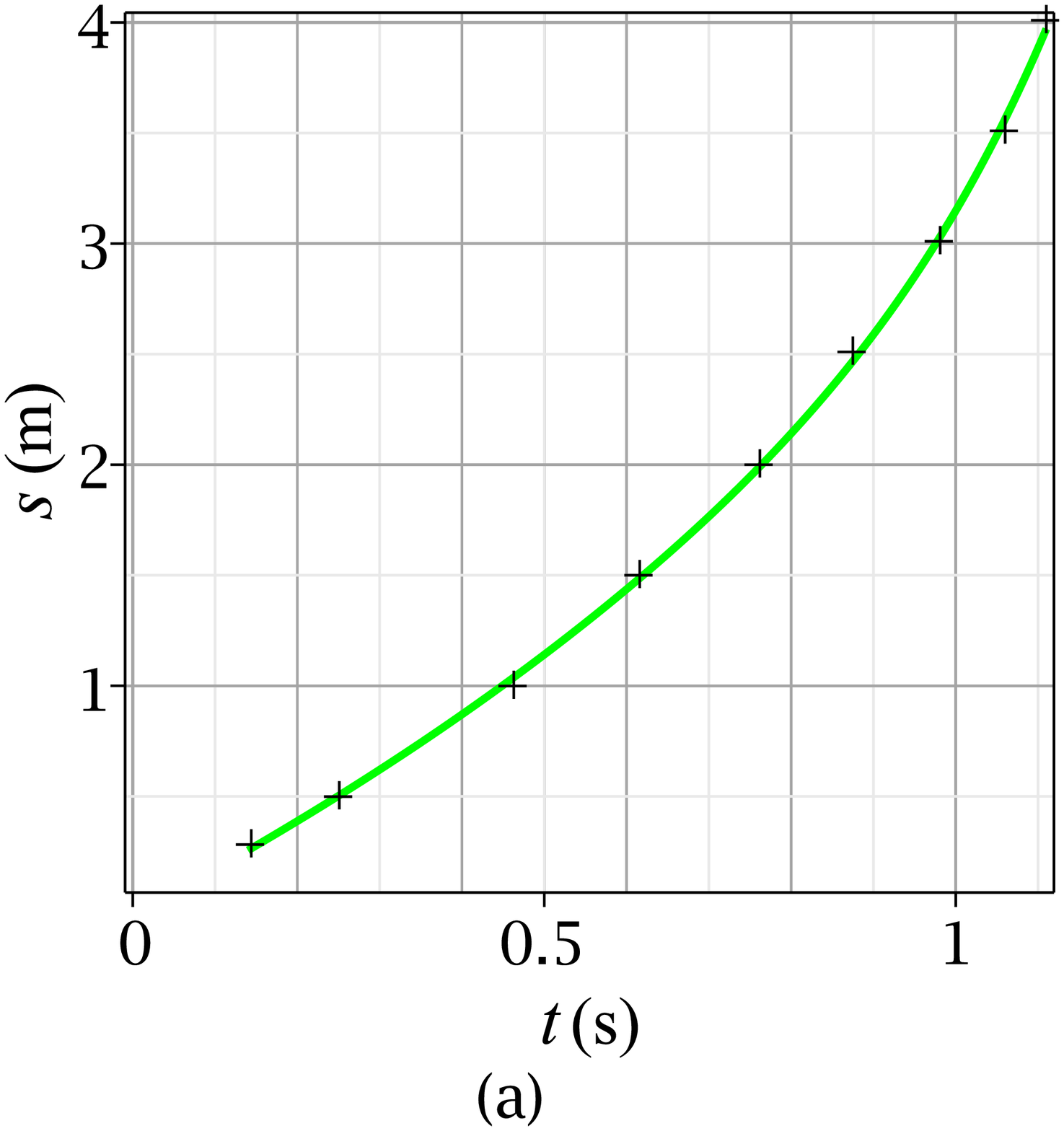}
\includegraphics[width=0.45\textwidth]{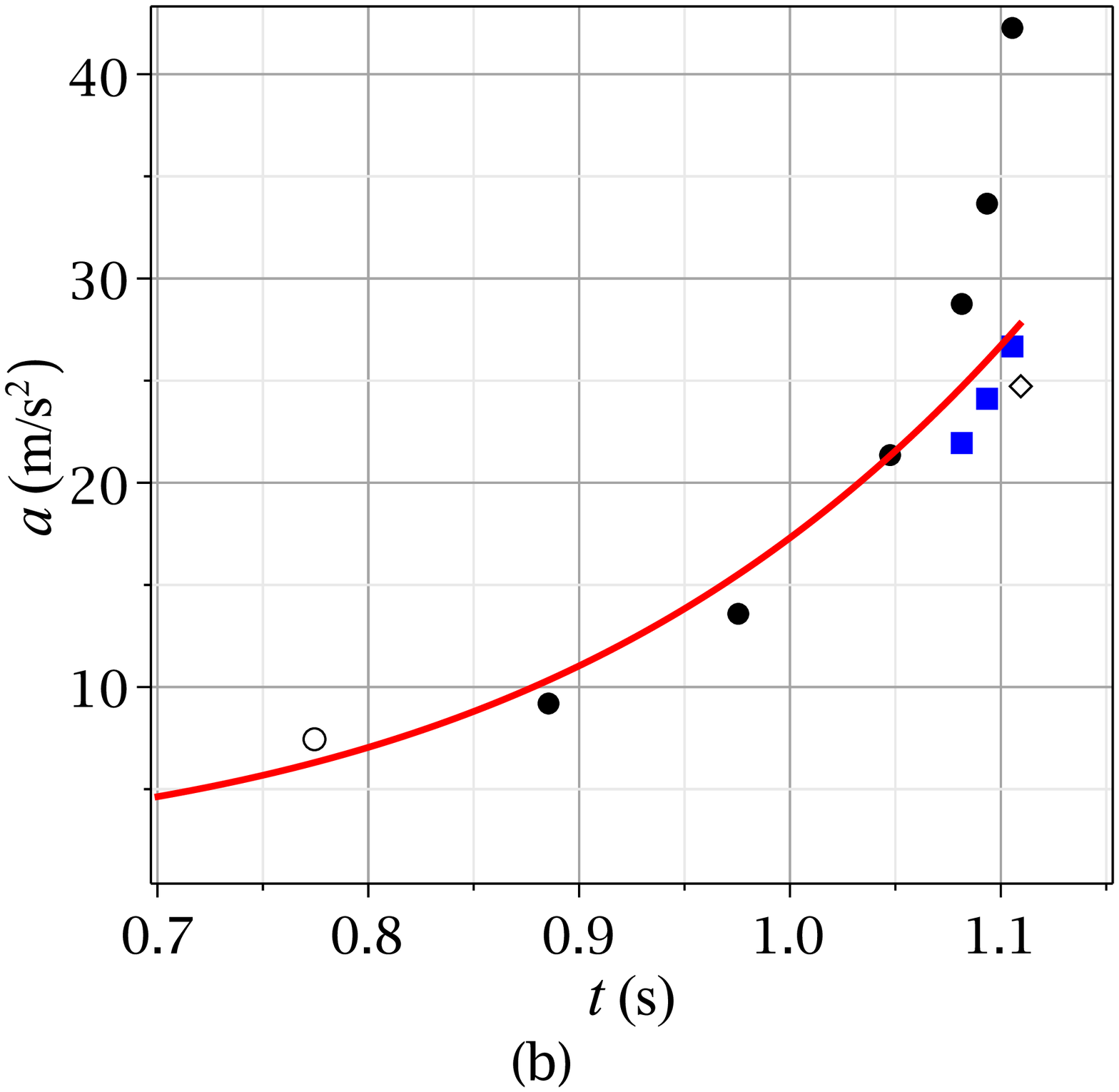}
\caption{(Color online) (a) Approximation of the $10\%$-methane flame trajectory from Fig.~\ref{fig8} by a polynomial $S(t) = -.037+2.03\,t+.3343\,t^2+.7856\,t^3-.3469\,t^4+.3829\,t^7.$ Same notation as in Fig.~\ref{fig9}. (b) Flame acceleration found as the second derivative of the polynomial $S(t)$ (solid line), and as  eigenvalues of asymmetric type IIa (squares), type IIb (circles), and symmetric type II (diamond) solutions. Open symbols denote metastable regimes.}\label{fig10}
\end{figure}

\begin{figure}
\includegraphics[width=0.6\textwidth]{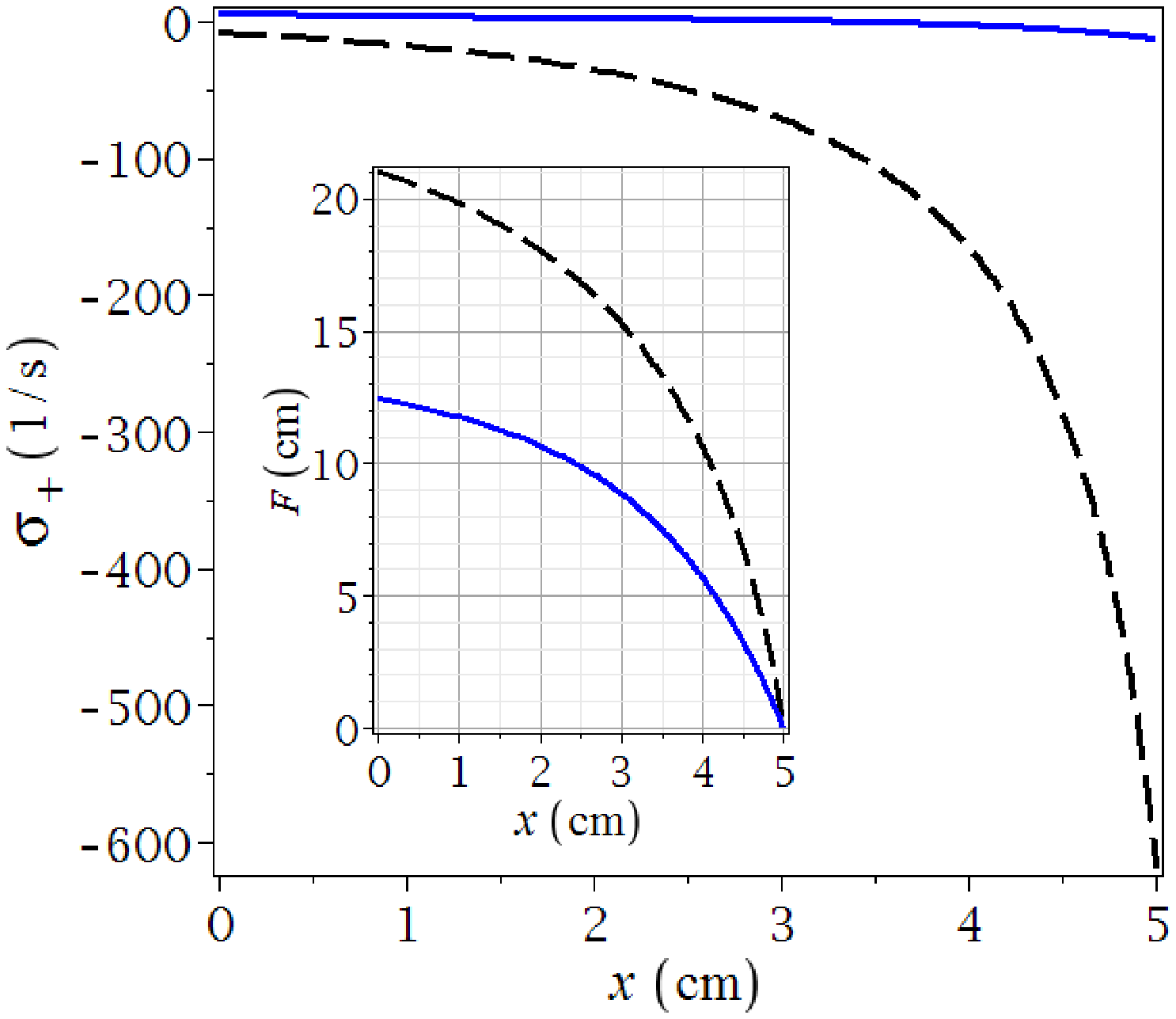}
\caption{(Color online) On-shell vorticity of the burnt gases and the front position (inset) as given by the type IIa (solid) and type IIb (dashed) solutions for a $10\%$-methane flame at $s=3.75$\,m. $\theta = 7.5,$ $U_f=40$\,cm/s. }\label{fig11}
\end{figure}

The other conclusive instance is the fastest $10\%$-methane flame. The heat losses are now negligible whatever flame symmetry because of the large normal speed. Moreover, calculations show that symmetric solutions appear only starting from $s\approx 4$\,m, that is at the very end of observation. On the other hand, this case is very interesting in that two different (asymmetric) type II solutions coexist in a vicinity of $s=3.8$\,m. As before, solutions with the lower (higher) value of $U$ will be refereed to as type IIa (type IIb). Asymmetric solutions start to appear at $s\approx2$\,m, and are unique for each $s\lesssim 3.7$\,m. The second solution emerges at still larger distances, and the two solutions coexist for $3.7$\,m $\lesssim s\lesssim 3.95$\,m. On this interval, the $U$-eigenvalue of the first (type IIb) solution and its acceleration rapidly grow, reaching the values $U=225$\,cm/s and $a=4220$\,cm/s$^2$ at $s=3.95$\,m, beyond which this solution ceases to exist. At the same time, the type IIa solution shows no such anomaly, its speed and acceleration gradually increase on the same interval, and become large only near the upper end (at $s=4.5$\,m, {\it e.g.,} $U=265$\,cm/s, $a=7650$\,cm/s$^2$). Theoretical and experimental results for the flame acceleration are compared in Fig.~\ref{fig10}. This comparison leads to the conclusion that the flame does travel quasi-steadily on the interval $2$\,m $\lesssim s\lesssim 3.8$\,m, but then undergoes a transition from type IIb to type IIa regime, which occurs well before the moment of disappearance of type IIb solutions. Evidently, the process of transition is essentially unsteady. This is the exceptional case mentioned earlier where the bound (\ref{quasisteady1}) is insufficient to guarantee flame steadiness. In fact, the dimensionless $U$-eigenvalue drops from about $4$ to $2.5$ over the small distance $\Delta s \approx 2,$ so that the first condition (\ref{quasisteady}) is strongly violated during the transition despite the flame acceleration remains bounded [the second condition (\ref{quasisteady}) is not sensibly affected because the dynamic pressure term in $a$ is still dominated by the gravitational contribution]. Recalling that the value of $U$ is directly proportional to the flame front length, it follows that the regular quasi-steady evolution during which the flame gradually increases its length is broken by a transition between the two type II regimes, which should show itself as a sudden contraction of the flame, Fig.~\ref{fig11}(inset). Its measured propagation speed ($V$) and acceleration are only weakly affected by this transition, but the gas velocity field is drastically changed, as is evident from Fig.~\ref{fig11} indicating a significant drop in the burnt gas vorticity.

Incidentally, the fact that steady solutions do not exist for $s \lesssim 2$\,m probably explains an observed irregularity in the behavior of near-stoichiometric methane-air flames. Namely, it is known that the upward propagation of these flames is rather unstable. In particular, their measured speeds show a significant scatter, and can be considered constant only approximately even when the upper tube end closed. This is in contrast to the horizontal case where all flames propagate with well-defined speeds before the onset of vibratory movement (the acoustic instability). Calculations also show that type II solutions become metastable before they disappear as $s$ decreases, {\it viz.,} the asymmetric solution at $s=2$\,m has $\dot{Q}=-6.6.$

\subsubsection{Spontaneous acceleration of methane-oxygen flames}

\begin{figure}
\includegraphics[width=0.4\textwidth]{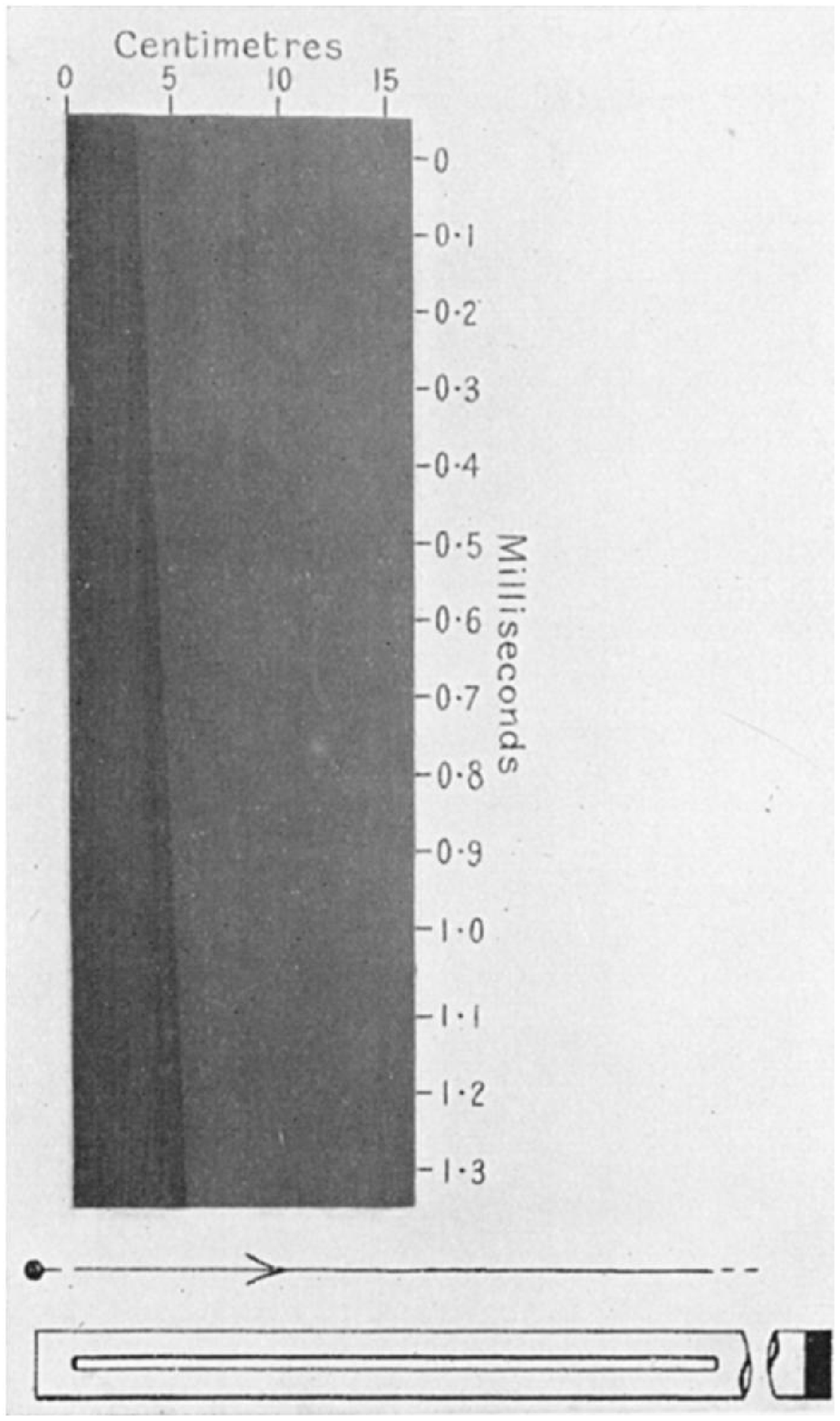}
\caption{Schlieren image of the initial segment of flame trajectory in a mixture of $33\%$ methane with oxygen, photographed onto a vertically moving film through a $30$\,cm long slit in the horizontal tube shown at the bottom. Flame propagates to the right. $b=2.5$\,cm, $L=1$\,m. {\it Source}: Fig.~3 of Ref.~\cite{payman1928} -- Reproduced by permission of The Royal Society of London.}\label{fig12}
\end{figure}

A quite different mechanism of flame acceleration acts at the earliest stage of propagation in a tube, before the compression waves produced by the ignition near its open end and reflected from the other end return to the flame. During this time interval, the longitudinal velocity of gases set into motion by the flame is determined by the hydrodynamic equations only up to an additive function of time. Indeed, the total kinetic energy of gases constitutes only a tiny fraction of the thermal energy released by deflagration. Therefore, in the approximation where this fraction is treated as negligible, the problem is degenerate in that any boost of the given solution describing the flame-induced flow yields another solution, the boost momentum being compensated by emission of backward compression waves. As usual, the rapid development of Darrieus-Landau instability (characterized by a time $\sim l_f/U_f$) slows down as the nonlinear interaction of unstable modes saturates. This interaction tends to form a quasi-steady curved flame, but the corresponding solutions now span a continuum as a result of the degeneracy. In view of the general tendency of flames to increase their propagation speed, it is natural to expect that the regime of uniform movement will be unstable against acceleration of the flame in the direction of its propagation. The flame will then evolve until the first stable regime is reached, which is the propagation with lowest acceleration sufficient to support steady curved flame. In the zero front-thickness approximation, the value of this spontaneous acceleration is, on dimensional grounds,
\begin{eqnarray}\label{spontaccel}
a = A\,\frac{U^2_f}{b}\,,
\end{eqnarray}
\noindent where $A$ is a numerical coefficient. It follows that this effect can be of interest only for flames with a sufficiently large normal speed; otherwise, the velocity gain is insignificant because duration of this phase of flame evolution is very short. It is well-known that the wall roughness as well as tube restrictions and enclosures promote flame acceleration and can trigger a deflagration-to-detonation transition. But the fact that sufficiently fast flames propagating from the open end of a tube with smooth walls can accelerate to very high speeds largely exceeding the normal speed was recognized already at the dawn of modern combustion science.\cite{dixon1903,payman1928} To quote H.~B.~Dixon, one of the pioneers of detonation studies, ``The propagation of the flame from the firing point is in most gaseous mixtures less rapid than the velocity of sound in the unburnt gas, but the rate of propagation of the flame augments, much more rapidly in some mixtures than in others. If the tube is a long one the flame will overtake the sound-wave after a more or less prolonged chase, according to the nature of the mixture.''\cite{dixon1903} Figure~\ref{fig12} is a Schlieren image\cite{payman1928} of a stoichiometric methane-oxygen flame propagating in a $2.5$\,cm diameter $1$\,m long iron tube. The tube was horizontal, but since the normal speed of the flame under consideration\cite{lecong2008,mazas2011} is very large, $U_f = (4.4\pm 0.3)$\,m/s, terrestrial gravity is negligible (Froude number is about $0.01$), so that the tube orientation is unimportant. Shown is a cut of the initial part of the flame travel from an open end, the other end being closed. Inspection of the image gives a value $V = 16.4$\,m/s for the mean laboratory flame speed on this segment. It takes about $0.005$\,s for the initial gas disturbance to reach the closed tube end and go back to the flame (the speed of sound in the fresh mixture $\approx 350$\,m/s). During this time interval, the flame travels a distance $s\approx 30$\,cm, being observed through a $30$\,cm long slit in the tube sketched at the bottom of Fig.~\ref{fig12}. The value of the flame speed reported\cite{footnote2} for this segment is $55$\,m/s.

To determine this value theoretically, one has to solve the system (\ref{master3}), (\ref{3rel}), (\ref{boundaryc1}) with $\theta = 10.2,$ $g\to a,$ and $a$ increased until the first stable regime is reached. Numerical integration shows that metastable ($\dot{Q}<0$) solutions appear first. $\dot{Q}$ increases with $a,$ and vanishes for $$A = 14.9.$$ This is for a symmetric flame; for asymmetric configuration, $A$ is twice as small. Putting this $A$ and $U_f=4.4$\,m/s into Eq.~(\ref{spontaccel}) gives $a=1.15\cdot 10^4$\,m/s$^2$ for a symmetric flame. The value of acceleration might grow further during subsequent flame evolution, but a simple estimate shows that its increase can be neglected on the interval under consideration. Indeed, the time scale characterizing the comparatively slow flame evolution after its stabilization, $b/U_f \approx 0.02$\,s, is significantly larger than its travel time, so that the flame acceleration over the distance $s=30$\,cm can be considered uniform. With $a=1.15\cdot 10^4$\,m/s$^2$ and an initial speed $16.4$\,m/s, the mean flame speed on this segment is $V = 50.5$\,m/s (whereas for an asymmetric flame, $V = 38.7$\,m/s). The measured speed is well within the calculational uncertainty of the found value.

\section{Discussion and conclusions}\label{consclusions}

The analysis carried out in this paper revealed existence of several steady regimes of upward flame propagation in a vertical tube. As in the horizontal tube, each of them belongs to one of the two families described by type I or type II solutions of the on-shell equations. In comparison with the horizontal case, the two continua of $U$'s parameterizing the type I and type II solutions are separated much wider: as the normal flame speed decreases, the ratio of the propagation speeds grows, reaching the values $5-7$ for near-limit flames. A fundamental qualitative distinction in the behavior of vertical and horizontal flames is that the former are more susceptible to the finite front-thickness effects. This distinction reflects an important difference in the small-$U_f$ asymptotics of the flame propagation speed in the two cases. Namely, while $U$ of a horizontal flame, calculated in the zero-thickness approximation, remains finite as $U_f\to 0,$ it turns out to be formally unbounded for a vertically propagating flame. By this reason, flame dynamics characterized by large Froude numbers is strongly affected by the coupling of the transport processes inside the front with the incoming flow strain: The large velocity gradients induced by the flame with a small $U_f$ give rise to appreciable corrections to its normal speed despite the fact that the Markstein length remains very small compared to the tube diameter. In particular, as discussed in Sec.~\ref{assessment}, an extremely large flame stretch characterizing the near-limit type II solutions must either change drastically the symmetric flame by reducing the front slope, or else preclude existence of configurations of this type. Since resolution of this issue is beyond the capabilities of the small-$l_f$ asymptotic expansion, experimental realization of the possibility mentioned in Sec.~\ref{reduction} is of special interest. An asymmetric type II near-limit flame can probably be obtained by means of a sufficiently strong distributed ignition, to surmount the large velocity gap separating type I and type II regimes.

On the other hand, the finite front-thickness effects turn out to be considerably weaker in type I regimes, and in most cases admit consistent treatment within the small-$l_f$ expansion. According to the results of Sec.~\ref{inflammability}, these effects produce the following trends in the behavior of near-limit flames:
\begin{itemize}
  \item A critical curve identifying the inflammability region boundary in the $U_f$--$\theta$ plane shifts rightwards (to larger $U_f$) as the Markstein length $\mathscr{L}_s$ increases.
  \item The propagation speed of critical flames decreases as $\mathscr{L}_s$ increases.
\end{itemize}
The latter observation implies that in spite of having larger $U_f$ and $\theta,$ a critical propane-air flame ($\mathscr{L}_s >0$) is slower than a critical methane-air flame ($\mathscr{L}_s <0$) in the same tube. This curious conclusion of the theory appears to find experimental support in Levy's experimental results discussed in Sec.~\ref{critical}. Although the difference in the measured flame speeds is comparable to the standard deviation, the experimental evidence is that in both $9.5$\,cm and $5.1$\,cm diameter tubes, the limit propane-air flames are slower on average than methane-air flames. This is the more striking that things ought to be exactly the opposite according to the seemingly more intuitive view on the limit flame propagation as a buoyancy-driven process (in this picture, the flame speed equals that of a rising bubble of hot gases,\cite{collins1966} $U \sim \sqrt{(1-1/\theta)gb}$\ ). However, as was already stressed, flame dynamics in a strong gravity is  governed not by buoyancy, but by an interaction of the baroclinic vorticity with that generated by the front curvature. As to the bubble analogy, one of the conclusions of Sec.~\ref{dynamics} is that a steady flame never behaves like a bubble, because the burnt gas velocity remains large compared to $\theta U_f$ everywhere across the tube. This analogy emerges only at the latest, essentially unsteady stage of flame evolution, in the course of flame extinction.

It follows from what has just been said and from the remark made in Sec.~\ref{remark} that in order to draw a definite conclusion as to the finite front-thickness effect on the flame propagation speed, it is necessary to
\begin{itemize}
  \item Carry out the flame speed measurements for a wider variety of mixtures with significantly different transport properties.
  \item Consistently determine the flow strain Markstein length in these mixtures, to permit a more accurate theoretical prediction of the flame speed.
\end{itemize}
But independently of the precise value of the finite front-thickness correction, the main conclusion of Sec.~\ref{inflammability} is that the ultimate cause of flame extinction in a vertical tube is purely hydrodynamical, {\it viz.}, the burnt gas slowdown near the tube centerline taking place in a strong gravity. The finite front-thickness effects may increase the local burning rate in the flame centre, as is the case with methane-air flames, or decrease it (propane-air flames), but the gas flow stoppage occurring as the result of transition into a supercritical type Ib regime disrupts the burning in the flame centre, which then rapidly spreads towards the walls.

Turning finally to the flame dynamics in open tubes, the results of Sec.~\ref{acceleraion} leave no doubt that as in the horizontal case,  flame acceleration observed in vertical tubes with smooth walls is largely apparent, that is associated with an acceleration of the medium in which the flame propagates. This acceleration occurs either because of a difference in the boundary conditions at the tube ends, or  spontaneously, at the early stage of flame propagation. As a matter of fact, calculations show that in tubes with $L\gg b,$ the rate of change of the $U$-eigenvalue, that is the flame acceleration with respect to the fresh mixture, is comparatively small over most of the tube, reaching appreciable values only near its upper end. An important exception is delivered by transitions between different regimes of flame propagation, such as the flame contraction discussed in Sec.~\ref{openaccel}. A direct experimental study of these transitions would help elucidate their impact on flame dynamics, and determine relative stability of various regimes of flame propagation.

\acknowledgments{This study was partially supported by RFBR, research project No. 13-02-91054~a.}


\begin{thebibliography}{}
\bibitem{footnote1}
In this respect, vertical flame propagation should be contrasted to the horizontal, where the law $U \sim \sqrt{gb}$ holds in all mixtures, provided that $U\gg U_f.$
\end{thebibliography}
\end{document}